\documentclass[12pt]{article}
\usepackage{epsf}

\newcommand{\be}{\begin{equation}}
\newcommand{\ee}{\end{equation}}
\newcommand{\bea}{\begin{eqnarray}}
\newcommand{\eea}{\end{eqnarray}}

\title{\begin{flushright}
{\small SU{-}4240{-}705}
\end{flushright} 
Interacting Topological Defects on Frozen Topographies}

\author{\small \\ Mark~J. Bowick$^{(1,2)}$\thanks{\tt 
bowick@physics.syr.edu} \, 
David~R. Nelson$^{(2)}$\thanks{\tt nelson@cmt.harvard.edu} and Alex
Travesset$^{(1)}$\thanks{\tt alex@suhep.phy.syr.edu}
\\ $^1$Physics Department, Syracuse University,\\
Syracuse, NY 13244-1130, USA 
\\ $^2$Lyman Laboratory of Physics, Harvard University, \\
Cambridge, MA 02138, USA \\ }

\date{}
\begin{document}
\maketitle
\begin{abstract}

We propose and analyze an effective free energy describing the physics
of disclination defects in particle arrays constrained to move on an
arbitrary two-dimensional surface. At finite temperature the physics
of interacting disclinations is mapped to a Laplacian Sine-Gordon 
Hamiltonian suitable for numerical simulations. We then specialize 
to the case of a spherical crystal at zero temperature.  
The ground state is analyzed as a function of the ratio of the defect 
core energy to the Young's modulus. We argue that the core energy
contribution becomes less and less important in the limit $R \gg a$,
where $R$ is the radius of the sphere and $a$ is the particle
spacing. For large core energies there are twelve disclinations 
forming an icosahedron. For intermediate core energies
unusual finite-length grain boundaries are preferred. 
The complicated regime of small core energies, appropriate to the
limit $R/a \rightarrow \infty$, is also addressed. 
Finally we discuss the application of our results to the classic 
Thomson problem of finding the ground state of electrons 
distributed on a two-sphere.

\end{abstract}
\section{Introduction}

The theory of two-dimensional melting of essentially planar materials
(monolayers) is a rich and well-developed subject \cite{KTNHY,REV1}. An
interesting aspect of melting in this low dimension is that both the
crystalline to hexatic and hexatic to fluid transitions can be driven by
the sequential liberation of point-like topological defects 
{--} dislocations in the former case and disclinations in the latter. 
It is clearly important, therefore, to have a 
thorough understanding of the statistical
mechanics of interacting topological defects. On the plane all
topological defects are bound at zero temperature, but on manifolds
with more complicated topology excess free disclinations must exist even at
zero temperature. 

The statistical mechanics of particles confined to frozen surfaces of
constant positive and negative curvature was discussed, e.g. in
references \cite{DRNNew1} and \cite{DRNNew2}. It was argued that
regions of positive and negative curvature would promote the formation
of unpaired disclinations, and that these might be screened by clouds
of dislocations. At low temperature, it was suggested that the
anisotropic interaction between these screening dislocations would
lead them to condense into grain boundaries. The physics of particles
on a quenched {\em random} topography was discussed in Ref.\cite{DRNNew3}.

The simplest example of a surface with positive Gaussian curvature is
the sphere. Dodgson studied the ground state of the Abrikosov flux
lattice in a model thin film superconductor on a sphere (subject to a
field radiating from a magnetic monopole at the center), and found
evidence for twelve five-fold disclination defects at the vertices of
an icosahedron in an otherwise six-coordinated crystalline environment
\cite{DOD}. This defect configuration is similar to one proposed by
Lubensky and collaborators for lipid bilayer vesicles in the hexatic
phase \cite{MacLub}, except that in hexatics the disclination energy
is reduced by screening due to an equilibrium concentration of unbound
dislocations. Later, Dodgson and Moore proposed adding dislocations to
the ground state of a sufficiently large vortex crystal in a spherical
geometry to screen out the strains associated with twelve extra
disclinations in the Abrikosov phase \cite{DodMoore}. Vortices in a
thin film superconductor behave like particles interacting with a
repulsive logarithmic pair potential. Another context in which
crystalline ground states on a sphere arise is the so-called {\em Thomson
problem}, where the vortices are replaced by particles interacting
with a repulsive $1/r$ potential \cite{JJT,THOM,NON_ICO_TH}. Our own
interest in this class of problems was stimulated by the beautiful work
of Alar Toomre \cite{ALAR}, which we discuss later (and which
hopefully will be described by Toomre himself one day!). Toomre's
ideas also play a key role in a recent paper on the Thomson problem by
P\'erez-Garrido and Moore \cite{MOPZ}. For a discussion of
disclination and dislocation defects for disk-like configurations of
electrons in the plane see \cite{KouShk}.

The study of melting and the nature of the ground state on curved
manifolds may be a good testing ground for understanding the new
features that arise from the topological defects required for particle
arrays on surfaces with nontrivial topologies. Our approach is to work
directly with the defects themselves, and treat the particles within
continuum elastic theory. This approach is more general than, say, a
direct simulation of particles interacting with a logarithmic or $1/r$ potential,
because all details of the pair potential are embodied in the elastic
constants mediating the interaction between defects. By eliminating
explicit reference to the particles themselves, we also greatly reduce
the number of degrees of freedom needed to study the ground state. As
we shall see, the effective Hamiltonian used here, in which defects
such as grain boundaries and dislocations are built up out of
elementary disclinations, leads to a variety of interesting and novel
structures not encountered in the plane. 

The statistical mechanics of monolayers on curved surfaces such as the
sphere may also be viewed as the infinite bending rigidity limit of
membranes with a spherical topology. Our investigation may therefore be
considered a prelude to the careful incorporation of defects in 
the study of the phase transitions of, e.g., membranes composed 
of lipid bilayers \cite{REV2}.

\begin{figure}[htb]
  \epsfxsize=3 in \centerline{\epsfbox{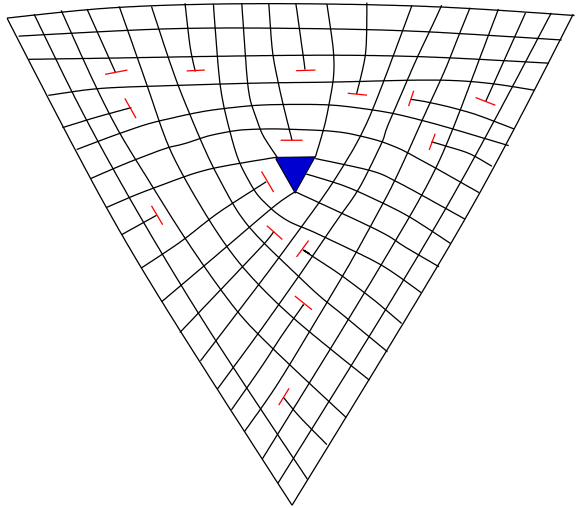}}
  \caption{Schematic of an isolated threefold disclination, in an
  approximately four-coordinated medium. The elastic stress in the
  vicinity of the isolated disclination is relieved by the
  formation of a screening cloud of dislocations.}
  \label{fig__extradiscl}
  \end{figure}

It is useful to review expectations for low
temperature configurations of crystals in flat space \cite{CN}.
Although the ground state is believed to be defect free, one can
certainly consider the response to adding a single excess
disclination. The stresses induced by such a disclination are very
high, and the energy can be lowered by
polarizing the surrounding medium into dislocation pairs, as indicated
schematically in Fig.~\ref{fig__extradiscl}.

When interactions between dislocations are taken into account one 
might expect them to organize into grain boundaries 
(i.e., lines of dislocations with
Burgers vectors oriented perpendicular to the lines) to minimize the
energy even further. Experiments on smectic liquid crystal films with
tilted molecules \cite{DPM} (the tilt is used to force in an extra
disclination) reveal a pattern of five jagged grain boundaries radiating
outward, consistent with this picture. Computer simulations with
periodic boundary conditions have been used to study the relaxation of
a disclination quartet (two fives and two sevens), from an initial
configuration where these defects sit on the corners of a very large
square in an otherwise six-coordinated medium \cite{SomCanKap}. 
After the relaxation, grain boundaries joining the fives to the sevens appear. 
Relaxation of the disclination elastic stresses in this way occurs 
at a price {--} the core energies associated with the extra
dislocations lead to an additional term in the energy which diverges
linearly with system size $R$, as compared to the $R^2$ divergence 
associated with an unscreened disclination \cite{CHUI}. 

A situation reminiscent of these flat space experiments occurs on 
surfaces of non-zero Gaussian curvature, e.g., the sphere. 
Although the Gaussian curvature of the
sphere approximately compensates the strains associated with isolated
disclinations, for small core energies (or, equivalently, large sphere
radius compared to the particle spacing) it can still be favorable to
introduce extra screening dislocations into the ground state.  

To see how screening of an isolated five-fold disclination by
dislocations comes out on a sphere, it is helpful to first consider
what happens in flat space. A five-fold disclination can be created by
removing a wedge of material subtending an angle $s=2\pi/6$ 
and then deforming the remaining material to close the gap. (The
disclination in the square lattice of Fig.\ref{fig__extradiscl} was
made by removing a $2\pi/4 = 90^{\circ}$ wedge). The resulting
stresses were calculated, e.g., in Ref.~\cite{SeuNel}. We use polar
coordinates $r$ and $\phi$, measured from the center of the
disclination. If $\mu$ and $\lambda$ are the material elastic
constants, the stress tensor is dominated by $\sigma_{\phi \phi}$,
where (neglecting logarithmic corrections due to boundary effects),
\be\label{stress}
\sigma_{\phi \phi} = \frac{K_0 s}{4\pi} \, ,
\ee
and $K_0$, the Young's modulus, is related to the Lam\'e coefficients by 
\be\label{YoungMod}
K_0 = \frac{4\mu(\mu + \lambda)}{2\mu + \lambda} \, .
\ee
Note that $\sigma_{\phi \phi}$ is proportional to the disclination
charge $s$. This approximately constant stress arises from the stretching of
material required to close the gap engendered by the missing wedge,
and leads to the $R^2$ divergence in disclination energy with system
size $R$ \cite{SeuNel}. Consider now the fate of a tightly bound
dislocation pair (the Burgers' vectors are equal and opposite with
$|\vec{b}| = a$) placed in the stress field of this disclination. The
stress $\sigma_{\phi \phi}$ creates a Peach-Kohler force which tries
to tear the dislocation pair apart \cite{BruHalZip}. We assume for
simplicity a purely radial separation $\Delta r$ between dislocations
with Burgers' vectors in the tangential direction. The energy of the
pair then consists of $2E_d$ ($E_d$ is the dislocation core energy), a
logarithmic binding energy and a linear Peach-Kohler term (proportional to
$\sigma_{\phi \phi}$) which tries to ``ionize'' the pair, similar to
the effect of an electric field on a charge dipole,
\be\label{Peach-Kohler}
E_{pair}(\Delta r) = 2E_d + \frac{K_0 b^2}{4\pi} {\rm ln} (\frac{\Delta
r}{a}) - \frac{K_{0} b}{4\pi} s(\Delta r) \, .
\ee
The energy can be lowered once $\Delta r$ exceeds 
$\Delta r^* \approx b/s \approx a$ and the pair separates 
\cite{Remark}. One of the liberated dislocations moves off to infinity
while the other remains to help screen the disclination. As more and
more dislocations are created in this way, the stress is reduced until
the dislocation density $n_d$ in an annulus of width $dr$ at radius
$r$ from the disclination is \cite{BruHalZip} 
\be\label{dislocdensity}
n_d(r) \approx (\frac{s}{2\pi}) \frac{1}{ra} \, .
\ee
Note that if these dislocations collapse to form a single linear grain
boundary radiating out from the disclination, the angular deficit $s$ is
related to the spacing $l$ between dislocations in the grain by
$s \approx a/l$. If the dislocations form $m$ grain
boundaries, the spacing will be $l \approx am/s$. In this paper we
shall study the cases $m=2$ and $m=5$. 

The analysis sketched above is easily adapted to the curved surface
of a sphere of radius $R$. The angular deficit $s$ associated with a
circuit around a dislocation in flat space can now be compensated by
the nonzero Gaussian curvature $1/R^2$. Let us assume that a five-fold
disclination is placed at the north pole of the sphere. We describe
the physics by geodesic polar coordinates $(r,\phi)$ about this point
with metric
\be\label{metric}
ds^2 = dr^2 + R^2 {\rm sin}^2(\frac{r}{R})d\phi^2 \, ,
\ee
and work in the limit $R \gg a$. We expect that the stress is
controlled by the {\em effective} disclination charge inside a
circuit at fixed geodesic distance $r$ from the disclination (see
Eq.(\ref{free_geom_2}) below), namely
\bea\label{deficit}
s_{eff}(r) &=& s - \int^{2\pi}_0 {\rm d}\phi \int^r_0 {\rm d}r'
\sqrt{g}\, K 
\nonumber \\
&=& s - \frac{2\pi}{R^2} \int^r_0 R\,{\rm sin}(\frac{r'}{R}) {\rm d}r' =
\frac{\pi}{3} - 4\pi {\rm sin}^2(\frac{r}{2R}) \, .
\eea
Note that $s_{eff}(r)$ decreases with increasing $r$.
In the limit of weak curvature $R \gg a$, we expect that the stress
formula (\ref{stress}) is replaced by
\be\label{curvedstress}
\sigma_{\phi \phi}(r) = K_0 \frac{s_{eff}(r)}{4\pi} \, ,
\ee
with a corresponding weakening of the Peach-Kohler force.
The reduction in the angular stress $\sigma_{\phi \phi}$ with
increasing geodesic distance from the disclination arises because the
stretching required to remove a wedge in flat space is reduced
according to the metric (\ref{metric}). We now expect the dislocations
in $m$ grain boundaries radiating from a disclination to exhibit a
{\em variable} spacing between dislocations,
\be\label{spacing}
l(r) \approx \frac{a m}{s_{eff}(r)} \, .
\ee
Note that the spacing diverges as $r \rightarrow r_c^{-}$, where 
\be\label{rcritical}
\frac{r_c}{R} \equiv \Theta_c = {\rm cos}^{-1}\frac{5}{6} 
= 33.56^{\circ} \, .
\ee
The angular jump $\Delta s(r) = s_{eff}/m$ across the grain boundaries
thus becomes smaller with increasing $r$ and these boundaries
eventually terminate when the dislocations become sufficiently
dilute. The total number of dislocations contained in the $m$ grain
boundaries is approximately
\be\label{notanotherequation}
N_d \sim \frac{R}{a} s({\bf x}) \, .
\ee 
Our calculations support this picture, and we find 
that the extra dislocations seem to form grain boundaries. 
Remarkably, and in contrast to flat space, these grain boundaries 
do indeed stop or start inside the crystalline medium. 
Our results also hint at a branching pattern of grain boundary 
networks (each radiating from a disclination), reminiscent of those 
found in Ref.~\cite{DPM}.     

The paper is organized as follows.
In section ~\ref{SECT__Melting} we develop a formalism whose basic
degrees of freedom are the defects themselves, rather than the
underlying interacting particles. The particles themselves are treated
within continuum elastic theory. As mentioned above, such a 
formalism has the advantage of reducing the number of 
degrees of freedom as well as being rather
universal in the sense that it applies to a broad 
class of interaction potentials. Varying the pair potential simply
corresponds to changing the elastic moduli and defect core energy of
the model. The model has the advantage that defects can move directly
to positions which minimize the energy, without the constraints
associated with disclination motion or dislocation climb in a
crystalline medium which would attend a particle simulation. 
Despite its simplicity, finite temperature statistical mechanics of
this model is still not amenable to a direct analytic solution. 
A duality mapping to an equivalent Laplacian Sine-Gordon model,
however, yields a model with short range interactions 
whose lattice version should be straightforward to
simulate numerically.  

In contrast, the limit of zero temperature may be treated analytically
and we turn to this in the next three sections of the paper. In
particular we discuss the ground state of a spherical crystal as a
function of defect core energy relative to the combination of elastic
constants (Young's modulus) which determines defect interactions at
large distances. 

In section~\ref{SECT__Ico} we shift our attention from defects
alone to the underlying lattice structure. We first discuss lattices
with icosahedral symmetry.  Our formalism applied to this case
predicts the range of core energies for which the lattice is unstable
to the formation of defects.  

An interesting application for our formalism is to the  
{\em Thomson} problem \cite{JJT,THOM}, discussed in
section~\ref{SECT__Thomson}. The predictions of our
approach are in agreement with existing results where 
comparisons are available.
A beautiful experimental realization of the Thomson problem is provided
by multi-electron bubbles trapped in liquid helium at low 
temperatures \cite{AL}.
Order in electrons confined by a positively charged capacitor plate to
a helium surface has been studied for many years. Except for capillary
wave deformations, crystallization proceeds in an essentially flat 
environment. At high electron densities, curvature is introduced via
an instability to a regular array of ``dimples'' in the helium
surface, each containing a million electrons or more. Upon increasing
the density of positive charge below the surface further by adding a
metallic tip to the anode, one can form completely submerged
multi-electron spherical bubbles. Typical bubbles contain $10^6{-}10^8$
electrons. The outward electrostatic repulsion of the electrons on the
inner surface of the helium bubble balances against the surface
tension of the helium interface to produce bubbles with diameters in
the range $10{-}100$ microns. Results for the Thomson problem have
implications for trapped multi-electron bubbles well below
the flat space freezing temperature. 
      
\section{Finite Temperature}
\label{SECT__Melting}
\subsection{Free Energy}

As our main interest lies in the study of defects on two-dimensional 
curved surfaces, we need a formalism that deals directly with the defect 
degrees of freedom themselves. A rigorous geometrical derivation of
the effective free
energy for the defects is given in \cite{US1}. An equivalent
derivation may also be given by integrating out the
phonon degrees of freedom from the elastic Hamiltonian \cite{NEL1},
with the appropriate modifications for a general distribution of
defects.  The energy of a two-dimensional crystal embedded in an
arbitrary frozen geometry described by a metric $g_{ij}({\bf x})$ 
is given by 
\bea\label{free_geom}
E&=&K_0 \int d^2 {\bf x}\sqrt{g({\bf x})} d^2 {\bf y} \sqrt{g({\bf y})}
(K({\bf x})-s({\bf x)})\left.\frac{1}{\Delta^2}\right|_{\bf{x y}}
(K({\bf y})-s({\bf y)})
\nonumber\\
&+&
K_A\int d^2 {\bf x}\sqrt{g({\bf x})} d^2 {\bf y} \sqrt{g({\bf y})}
(K({\bf x})-s({\bf x)})\left.\frac{1}{\Delta}\right|_{\bf{x y}}
(K({\bf y})-s({\bf y)}) 
\eea
where $g({\bf x})$ is the determinant of the metric tensor,
$K({\bf x})$ is the associated Gaussian curvature and $s({\bf x})$ the
disclination density
\be\label{free_discl}
s({\bf x})=\frac{\pi}{3\sqrt{g({\bf x})}} 
\sum_{i=1}^N q_i \delta({\bf x},{\bf x}_i) \ ,
\ee
with $N$ disclinations located at the sites ${\bf x}_i$ of
an underlying triangulated particle array. 
The ``charges'' $q_i$ may be positive or negative.
Although we do not restrict the allowed values of the charge, we expect
the unit charge defects to dominate for energetic reasons. 
A plus one charge corresponds to a five-fold coordinated particle (a five-disclination)
and a minus one charge corresponds to a seven-fold coordinated particle (a
seven-disclination). Charges are attracted to regions of {\em like-sign}
Gaussian curvature. 

The first term of Eq.~(\ref{free_geom}) represents a long range elastic
interaction and $K_0$ is the Young's modulus of Eq.~(\ref{YoungMod}) \cite{NEL1}.

The second term in Eq.~(\ref{free_geom}) contains
a single inverse-Laplacian operator, which is singular at short distances
due to distortions of the lattice at distances less than the
lattice spacing. This is the dominant term for hexatic membranes,
where $K_A$ is the hexatic stiffness.\cite{NEL1} In the present
context, this singular contribution leads to a renormalized
core energy $E_{core}(K_A)$ for each defect and it represents
non-universal details of the interaction on the scale of the
inter-particle spacing $a$.  
The energy of Eq.~(\ref{free_geom}) is thus simplified to 
\bea\label{free_geom_2}
E(K_0)&=&K_0\int d^2 {\bf x}\sqrt{g({\bf x})} d^2 {\bf y} \sqrt{g({\bf y})}
(K({\bf x})-s({\bf x)})\left.\frac{1}{\Delta^2}\right|_{\bf{x y}}
(K({\bf y})-s({\bf y)})
\nonumber\\
&+&
N \, E_{core} \ .
\eea
Although it is not essential, we assume for convenience that the 
core energies of five and seven-fold disclinations are identical. 
The partition function of our model is then
\be\label{part_func}
{\cal Z}(\beta)=\sum_{N_+,N_{-}} 
\frac{\delta_{N_+-N_{-},6 \chi}}{N_+! N_{-}!}y^{N_++N_{-}}
\int \prod_{\mu=1}^{N_+}d {\bf x}_{\mu}^+\sqrt{g}
\prod_{\nu=1}^{N_{-}}d {\bf x}_{\nu}^{-}\sqrt{g} e^{-\beta E(K_0)} \ ,
\ee
where $E(K_0)$ is the first term in Eq.~(\ref{free_geom_2}),
$y$ is the disclination fugacity $e^{-\beta E_{\rm core}}$ ($\beta$ is
the inverse temperature), $N_{\pm}$ 
is the total number of fives and sevens respectively and $\chi$ is the
Euler characteristic of the surface.  
For a given microscopic interaction potential both $y$ and $K_0$ are
fixed. We shall find it useful, however, to regard these as independent
parameters and discuss, in particular, the limits of large and small
$E_{core}$ compared to $K_0 a^2$, where $a$ is the lattice constant. 

Despite its elegant form this model is difficult to solve
analytically. It is, moreover, challenging for direct numerical
simulation because of the long-range interaction embodied in
$1/\Delta^2$ {--} see the explicit form for $E(K_0)$ given in
Eqs.~(\ref{bi_harm_sol}) and (\ref{energy_cosb}) below.  An
alternative formulation is suggested by the Laplacian roughening model
for flat space melting \cite{NEL2,STR}. 
Direct molecular dynamics simulations or energy evaluations of 
particles interacting with a specified potential
\cite{PGM,ALAR} are also of considerable interest. Since this approach takes
the particles as the primary degrees of freedom, rather than the
defects, it falls outside the scope of the present paper.  

\subsection{The Sine-Gordon model}\label{sinegordon}

We now restrict ourselves to the case of the sphere, which has Euler
characteristic $\chi=2$. We map the previous model to a dual 
Sine-Gordon model with only short range interactions by adapting the
derivation presented in \cite{LUB1} to the present case, with some
additional improvements.

Let us start with the identity
\bea\label{ident_def}
&&e^{-\frac{\beta}{2}\int\sqrt{g} du \sqrt{g} dv 
(s(u)-K(u))\frac{1}{\Delta^2}(s(v)-K(v))}
\nonumber
\\
&=&(det'\Delta^2)\int {\cal D} \phi' 
e^{-\frac{1}{2\beta}\int d u \sqrt{g} \Delta \phi \Delta \phi}  
e^{-i\int d u \sqrt{g} \phi(u) (s(u)-K(u))} 
\eea
The topological constraints ensure that the zero mode does not
contribute to the path integral and this is indicated by the primes in
the determinant and the measure. 
 
Since the zero mode is the constant eigenvector of the Laplacian,
orthonormality implies that
\be\label{phi_orthonormal}
\int du \sqrt{g} \phi=0 
\ee
for any configuration $\phi$ included in the measure of the path
integral in Eq.~(\ref{ident_def}). 

With this identity Eq.~(\ref{part_func}) becomes 
\be\label{int_part_func} {\cal Z}(\beta)=\int {\cal D} \phi'
{\cal F}(\phi) e^{-\frac{1}{2\beta}\int du \sqrt{g} \Delta \phi \Delta
\phi} e^{-i\int d u \sqrt{g} \phi K(u)}\equiv\int {\cal D} \phi'
e^{-{\cal H}(\phi)} \ , 
\ee 
where the last identification defines the Hamiltonian ${\cal
H}({\phi})$ and ${\cal F}(\phi)$ is given by
\bea\label{new_part_func} 
{\cal F}&=&\sum_{N_{+},N_{-}} \frac{\delta_{N_+-N_{-},12}}
{N_{+}!N_{-}!}y^{N_++N_-} \int \prod_{\mu=1}^{N_+} d {\bf
x}^+_{\mu}\sqrt{g} \prod_{\nu=1}^{N_-} d {\bf x}_{\nu}^{-} \sqrt{g}
e^{\frac{i\pi}{3}( \phi({\bf x}^+_{\mu})-\phi({\bf x}_{\nu}^{-}))}
\nonumber\\ 
&=&
\sum_{N_{+},N_{-}} \frac{\delta_{N_+-N_{-},12}} {N_{+}
N_{-}}(y\int du \sqrt{g} e^{\frac{i\pi}{3}i\phi(u)})^{N_+} (y \int du
\sqrt{g} e^{\frac{-i\pi}{3}i\phi(u)})^{N_{-}} 
\eea 
Upon writing the Kronecker delta as  
\be\label{delta_identity}
\delta_{N_+-N_{-},12}=\int^1_0 dx \ e^{-i2\pi x(N_+-N_{-}-12)} \ , 
\ee
one finds 
\be\label{more_part} {\cal F}(\phi)= \int^1_0 dx \ e^{-i12x} e^{\int du 
\sqrt{g} \cos(\frac{\pi}{3}\phi(u)+x)} 
\ee 
Inserting this result in Eq.~(\ref{int_part_func}) and performing the integral over
the $x$ variable leads to 
\be\label{middle_expression} 
{\cal H}(\phi)=\frac{1}{2\beta}(\frac{3}{\pi})^2\int du \sqrt{g}
\Delta \phi\Delta\phi-2y\int du \sqrt{g} \cos \phi+
\frac{3i}{\pi} \int  du \sqrt{g} K \phi \ .  
\ee 
The last (imaginary) term is a nuisance for practical applications.  For
the case of the sphere, however, the Gaussian curvature is constant,
and we have 
\be\label{ugly_term} 
\int du \sqrt{g} K \phi=K \int du \sqrt{g} \phi=0 \ , 
\ee 
where we have used Eq.~(\ref{phi_orthonormal}).  
The Sine-Gordon representation for the sphere takes a very simple form 
\be\label{final_expression} {\cal H}(\phi)=
\frac{1}{2\beta}(\frac{3}{\pi})^2\int du \sqrt{g}
\Delta \phi\Delta\phi-2y\int du \sqrt{g} \cos \phi \ .  
\ee
Discretizing this expression for large $y$ will yield a simple model with integer
variables $\phi(u)$. A numerical simulation of this model seems the appropriate
way to study the finite temperature statistical mechanics of defect
arrays on a sphere \cite{NEL2,STR}.

We now turn to the limiting case of zero temperature.

\section{Zero Temperature Limit}\label{SECT__zerot}

\subsection{General Surfaces}

The zero temperature limit requires the determination of the ground
state by a minimization of the energy as a function of both the
position and total number of defects.

For the minimization with respect to the location of defects we see
that the energy Eq.~(\ref{free_geom_2}) depends only on the difference between
the geometric curvature and the defect density. As a result the
defects will arrange themselves to approximately match the Gaussian curvature
determined by the geometry of the confining surface. A complete screening of
the Gaussian curvature would yield a crystal with zero elastic energy at zero
temperature. An important example is that of a crystal with the
symmetry of a perfect icosahedron. The twelve positive disclinations
located at its twelve vertices compensate the Gaussian curvature. 
There are twelve five-fold coordinated particles at the vertices, 
and all the rest are six-fold coordinated.

As for the minimization with respect to total defect number, it is clear
that the second term of Eq.~(\ref{free_geom_2}) is linear with the
number of defects, and so will clearly favor the lowest
possible number of them.
The physics of the zero temperature limit is therefore controlled by
the competition between the core energy cost of creating a defect and
the compensating gain from the screening of Gaussian curvature when
defects are allowed to proliferate.

\subsection{The spherical crystal}\label{sub_sect_spher}

From now on we concentrate on a spherical crystal. 
Since the sphere has Euler characteristic 2 (genus 0) the charges
$q_i$ of a set of disclinations must satisfy 
\be\label{constr_sphere}
\int d^2 {\bf x}\sqrt{g({\bf x})} s({\bf x})=4\pi 
\rightarrow \sum_{i=1}^N q_i=12 \ .
\ee
This implies that, even at zero temperature, a sphere contains at least twelve 
excess five-fold disclinations.

\begin{figure}[htb]
  \epsfxsize=3 in \centerline{\epsfbox{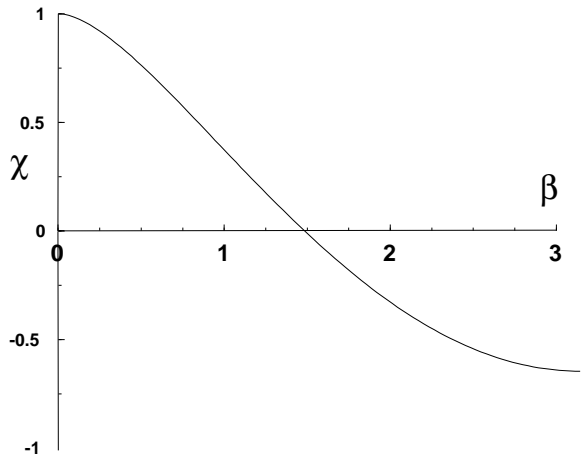}}
  \caption{Plot of $\chi/R^2$ as a function of the geodesic angle
  $\beta$. Only the interval $\beta \in [0,\pi]$ is plotted.}
  \label{fig__chi}
  \end{figure}

To evaluate the free energy Eq.~(\ref{free_geom}) we compute
first the inverse square-Laplacian operator on a sphere of radius $R$,
\be\label{bi_harm}
\frac{1}{4\pi}\chi(\theta^a,\phi^a;\theta^b,\phi^b)=\frac{1}{\Delta^2}=
R^2 \sum_{l=1}^{\infty}\sum_{m=-l}^l \frac{Y^l_m(\theta^a,\phi^a) 
Y^{l\ast}_m(\theta^b,\phi^b)}{l^2(l+1)^2} \ ,
\ee
where $Y^{l}_m(\theta,\phi)$ are the spherical harmonics and
$(\theta,\phi)$ are the usual spherical angles.  The $l=0$ term does not appear
in the sum, again as a result of the precise topology of the sphere 
(Eq.~(\ref{constr_sphere})). The absence of this zero mode leads to a
finite sum.  The expression Eq.~(\ref{bi_harm}) may also be written 
\be\label{bi_harm_Leg} 
\chi(\theta^a,\phi^a;\theta^b,\phi^b)\equiv\chi(\beta)
=R^2\sum_{l=1}^{\infty} \frac{2l+1}{l^2(l+1)^2} P_{l}(\cos \beta) \ ,
\ee 
where 
\be\label{cosbeta} 
\cos \beta=\cos
\theta^a\cos\theta^b+\sin\theta^a\sin\theta^b \cos(\phi^a-\phi^b) 
\ee 
gives the length $\beta$ of the geodesic arc connecting $(\theta^a,\phi^a)$
and $(\theta^b,\phi^b)$ on the sphere.
It is shown in Appendix \ref{SECT__App__sum} that this last
sum may be written \cite{Comparison}
\be\label{bi_harm_sol}
\chi(\theta^a,\phi^a;\theta^b,\phi^b)=R^2 \left( 1 +
\int^{\frac{1-cos\beta}{2}}_0 dz \, \frac{\ln z}{1-z} \right) \ .  
\ee 
In Appendix \ref{SECT__App__scal} we discuss the flat space limit of
infinite sphere radius. In Fig.~\ref{fig__chi} we plot $\chi/R^2$ (Eq.~(\ref{bi_harm_sol})) 
as a function of the geodesic distance
$\beta$. Although the formula Eq.~(\ref{bi_harm_sol}) is simple, it is
not particularly suitable for rapid numerical evaluation. 
In Appendix~\ref{SECT__App__Chi} we give alternative expressions for
$\chi$ better suited to fast numerical evaluation.

The final expression for the total energy of a spherical crystal 
with an arbitrary number of disclinations follows from
Eq.~(\ref{bi_harm_sol}) and Eq.~(\ref{free_geom_2}):
\be\label{energy_cosb}
E(K_0)=\frac{\pi K_0}{36} R^2 \sum_{i=1}^N\sum_{j=1}^{N} q_i q_j 
\chi(\theta^i,\phi^i;\theta^j,\phi^j)+N \, E_{core} \ .
\ee

Our interpretation of the disclination density screening out the Gaussian
curvature can be made more precise. Note that
\bea\label{discl_delta}
s({\bf x})&=&\frac{\pi}{3\sqrt{g}} \sum_{i=1}^N q_i \delta({\bf x},{\bf x}_i)
\nonumber\\
&=&\frac{1}{R^2}+\frac{\pi}{3 R^2}\sum_{l=1}^{\infty}
\sum_{m=-l}^l Y^l_m(\theta,\phi) 
\sum_{i=1}^N q_i Y^{l \ast}_m(\theta_i,\phi_i)
\\\nonumber
&=& K({\bf x})+\frac{\pi}{3 R^2}\sum_{l=1}^{\infty} \sum_{m=-l}^l 
Y^l_m(\theta,\phi) \sum_{i=1}^N q_i Y^{l \ast}_m(\theta_i,\phi_i) ,
\eea
where the topological constraint Eq.~(\ref{constr_sphere}) has been used.
This last identity makes it clear that the set of equations
\be\label{zero_screen}
\sum_{i=1}^N q_i Y^{l}_m(\theta_i,\phi_i)=0 \ , 
\ee
for all $l \geq 1$ and all $m$, is the condition that the disclination
density exactly matches the Gaussian curvature. Because it is
difficult to imagine how discrete disclination charges could exactly
cancel a smooth background Gaussian curvature, we cannot expect that
Eq.(\ref{zero_screen}) will be satisfied in general for all values of
$l$. We will, however, give examples where this set of equations is 
partially satisfied. It is easy to see, in fact, that in the limit of 
vanishing core energies, a configuration of defects satisfying 
Eq.~(\ref{zero_screen}) is an absolute minimum of the energy 
Eq.~(\ref{energy_cosb}), since the latter can be rewritten as
\be\label{energy_harm}
E=\frac{\pi^2K_0}{9} R^2 \sum_{l=1}^{\infty} \sum_{m=-l}^{l}
\frac{\left| \sum_{i=1}^N q_i Y^{l}_m(\theta_i,\phi_i) \right|^2}
{l^2(l+1)^2} +N\, E_{core}\ .
\ee
Eq.~(\ref{zero_screen}) then implies, for $E_{core}=0$, that the
energy attains its minimum value of $0$.

Finally, we note that an equivalent expression for
Eq.~(\ref{energy_cosb}) is given by
\be\label{energy_Leg}
E=\frac{\pi K_0}{36} R^2  \sum_{l=1}^{\infty} \frac{2l+1}{l^2(l+1)^2} \sum_{i=1}^N
\sum_{j=1}^N q_i q_j P_{l}(\cos \beta_{ij}) +N \, E_{core} \ .
\ee
Eqs.~(\ref{energy_harm}) and (\ref{energy_cosb}) are useful because they
express the total energy as a sum
of individual $l$-mode contributions
\be\label{energy_suml}
E=\sum_{l=1}^{\infty} E_l \ ,
\ee
with the order of magnitude of each $l$-mode coefficient being roughly
\be\label{rough_cont}
E_l \sim \frac{2l+1}{l^2(l+1)^2} \ .
\ee
By considering increasingly exotic arrangements of defects, we might
hope to satisfy Eq.~(\ref{zero_screen}) for more and more low $l$-modes.
If we do not enhance the large-$l$ contributions and do not pay too
large a price in defect core energy, then the total energy will be
small. 

\section{Large core energies: The icosahedral lattice}

In the limit of large core energies the creation of additional defects
will be strongly penalized and the sphere will contain only the minimum
allowed twelve positive disclinations. From symmetry considerations
it is a good ansatz to assume that these twelve disclinations minimize the
repulsive $\chi$ potential acting between them by forming an
icosahedron ${\cal I}$ \cite{DOD,MacLub}. It is not difficult to
check that the icosahedron is in fact an extremum of the energy
Eq.~(\ref{energy_cosb})
\be\label{der_icos} \left.\frac{\partial
E}{\partial \theta_i}\right|_{{\cal I}}=0 \ \ , \ \
\left.\frac{\partial E}{\partial \phi_i}\right|_{{\cal I}}=0 \ ,
\ee
where $i=1,\cdots,12$. We have checked numerically that fluctuations
around this extremum increase the energy. Allowing the fluctuations to
relax results in fast convergence to the icosahedron. Our numerical
minimization gives the icosahedron as a global minimum.  Thus our
model successfully predicts an icosahedron minimum in the case where
just twelve disclinations are allowed. 

From Eq.~(\ref{energy_harm}) the energy is a function of the quantity 
\be\label{triv_rep}
V^l_m({\cal I})=\sum_{i=1}^{12} Y^{l}_m(\theta_i,\phi_i) 
\ , \ \ m=-l,\cdots,l
\ee
where $(\theta_i,\phi_i)_{i=1,\cdots,12}$ are particular coordinates
for an icosahedron ${\cal I}$ on the sphere. This solution is obviously
invariant under the full icosahedral group plus inversions, 
${\cal Y}_h={\cal Y}\times C_i$.
Since ${\cal Y}$ is contained in $SO(3)$, we can construct a representation
of ${\cal Y}$ out of the irreducible representations of $SO(3)$. We have
\be\label{icos_act}
\sum_{m'=-l}^l D^l_{m m'}(g_Y)V^l_{m'}= 1\cdot V^l_m \ ,
\ee
where $g_Y$ is any element belonging to ${\cal Y}$.  That is, $V^l_m$
is a singlet of the icosahedral group ${\cal Y}$. This in turn means
that if the trivial representation (the so-called $A$ representation)
of the icosahedral group is not contained as an induced representation
from the full rotational group, then  Eq.(\ref{zero_screen}) is exactly satisfied
for any twelve disclinations forming an icosahedron,
\be\label{triv_rep_set0} V^l_m=\sum_{i} Y^{l}_m(\theta_i,\phi_i)=0 \ ,
\ \ m=-l,\cdots,l \, . 
\ee 
It remains to identify those values of $l$ which contain the trivial
representation. This is easily answered from an
analysis of the characters of the group. The number of trivial
representations $n_A$ contained in the $l-$th representation of
$SO(3)$ is given by 
\bea\label{triv_so}
n_A(l)&=&\frac{1}{60}\left\{2l+1+12\frac{\sin\left\{(l+\frac{1}{2})\frac{2\pi}{5}\right\}}
{\sin(\frac{\pi}{5})}+12\frac{\sin\left\{(l+\frac{1}{2})\frac{4\pi}{5}\right\}}
{\sin(\frac{2\pi}{5})}+\right.  \nonumber\\ && \left.
20\frac{\sin\left\{(l+\frac{1}{2})\frac{2\pi}{3}\right\}}
{\sin(\frac{\pi}{3})}+15\sin\left\{(l+\frac{1}{2})\pi\right\} \right \} \ , 
\eea
which is nonzero for $l=6,10,12,16$ and all even-$l$, $l>16$ \cite{SNR}.  
Note that Eq.~(\ref{triv_rep_set0}) is satisfied for all $l$-odd
modes, as follows from applying the inversion operator $I$, the generator
of the $C_i$ subgroup of ${\cal Y}_h$.

The icosahedral solution screens out the Gaussian curvature very
effectively. Eq.~(\ref{zero_screen}) is partially satisfied,
particularly for low $l$. The icosahedral lattice
allows for non-zero contributions for only three ($l=6,10,12$) of the first 
fifteen putative contributions in Eq.~(\ref{discl_delta}). A numerical
evaluation gives the energy of an icosahedron $E^{\cal I}$ as
\be\label{energy_icos_ex}
E^{\cal I} = 0.604\,(\frac{\pi K_0}{36} R^2) + 12\,E_{core} \ ,
\ee
where $E_{core}$ is the core energy of a single
disclination. Its precise value is non-universal and depends on 
short distance details of the microscopic pair potential.  
The coefficient of $\pi K_0 R^2/36$ is
universal, independent of short-distance properties. Let us study it
in more detail. In Table~\ref{tab__icos_fun_l} we show the relative
contribution from each $l$-mode. It is apparent that the first allowed
non-zero contribution $l=6$ accounts for almost $80\%$ of the total
energy of the icosahedron. Note also the relatively rapid convergence
of the expansion; truncating up to the $l=100$ mode gives a result
which differs by less than $0.2\%$ from the actual result,
Eq.~(\ref{energy_icos_ex}). It is remarkable how much the energy would
be reduced by canceling out the $l=6$ mode without further enhancing
higher $l$-modes.

  \begin{table}[htb]
  \centerline{
  \begin{tabular}{|c|l|l||c|l|l|}
  \multicolumn{1}{c}{$l$}   &
  \multicolumn{1}{c}{$E_l^{\cal I}$} & \multicolumn{1}{c}{$E^{\cal I}_t$} &
  \multicolumn{1}{c}{$l$}   &
  \multicolumn{1}{c}{$E_l^{\cal I}$} &  \multicolumn{1}{c}{$E^{\cal I}_t$} 
  \\\hline
   6  & $0.4669$  & $0.4669$ & $30$    & $0.0017          $ & $0.5925$ \\\hline
  10  & $0.0329$  & $0.4999$ & $40$    & $0.0012          $ & $0.5975$ \\\hline
  12  & $0.0507$  & $0.5506$ & $50$    & $0.0004          $ & $0.5997$ \\\hline
  16  & $0.0129$  & $0.5635$ & $80$    & $0.0001          $ & $0.6025$ \\\hline
  18  & $0.0125$  & $0.5760$ & $100$   & $3\times10^{-5}  $ & $0.6031$ \\\hline
  20  & $0.0004$  & $0.5764$ & $\infty$& $0               $ & $0.6043$ \\\hline
  \end{tabular}}
  \caption{The first column is the particular mode considered. $E^{\cal I}_l$ 
  is the contribution of the $l$-mode to the total energy and $E^{\cal I}_t$
  the running sum after adding all modes less than or equal to $l$. For
  convenience we set $E_{core}=0$ and $\pi K_0 R^2/36=1$ in this table.}
  \label{tab__icos_fun_l}
  \end{table}

\section{Small core energies: The proliferation of defects}\label{SECT__SmallCore}

If the defect core energies are small then the elastic energy may be reduced
by creating additional defects. The 
topological constraint Eq.~(\ref{constr_sphere}) requires
that additional defects appear in pairs of opposite charge. 
The challenge now is to understand and study the different possible 
distributions of these charges and the reduction in energy that 
those bring about when compared with the pure icosahedral case.
The general form of the energy is, similarly to 
Eq.~(\ref{energy_icos_ex}),
\be\label{energy_general}
E = C\, \frac{\pi K_0 a^2}{36}\left( \frac{R}{a}\right)^2 + N \, E_{core} \ ,
\ee
where we introduce the $C$-coefficient as a convenient parametrization
of the elastic part of the energy. In the limit $R/a \rightarrow
\infty$, where $a$ is the particle spacing, we expect grain boundaries
containing $N \sim R/a$ dislocations emerging from each
disclination. Hence the elastic term will {\em always} dominate over
the core energy term in this limit. The critical sphere radius $R=R_c$
above which long range elastic energies dominate is given by $R_c \sim
{\rm const.} (36 E_{core}/\pi K_0 a^2)a$.    

If the total number of defects is large, an unconstrained minimization of
Eq.~(\ref{energy_cosb}) becomes an involved numerical problem. Instead
of pursuing this further, we develop different approximations that
allow us to tackle the case of a large number of defects while still
capturing the most important features of the problem.

With results for a particular defect ansatz expressed as in
Eq.~(\ref{energy_general}), we need to determine the minimum distance of
closest approach of neighboring plus-minus defect pairs. Upon
identifying this distance with $a$, we can obtain
the total number of particles $M$ embodied in the defect configuration
via the identification 
\be\label{partno}
M \approx \frac{8\pi}{\sqrt{3}a^2} R^2 \, .
\ee

\subsection{The icosahedral approximation}

Let us add new sets of twelve defects, each set lying on the vertices of an
icosahedron. That is, we consider Eq.~(\ref{energy_cosb}), not as a
function of individual defects, but as a function of icosahedra of
defects. From the mathematical arguments in the previous section, we
can guarantee that the $l$-modes which vanish in the expansion
Eq.~(\ref{energy_harm}) for the pure icosahedral case, will continue to
do so within this approach. Since most of the low $l$-modes, which
dominate the energy, vanish for any icosahedron, we expect that the
Euler angles of the sets of icosahedra may be arranged to cancel the
remaining non-vanishing low-$l$ contributions. Our hope is that the 
energy bounds derived from this constrained problem provide a 
reasonable picture of the full unconstrained model.

If there are $n_{+}^{\cal I}$ icosahedra of fives, and $n_{-}^{\cal I}$ 
icosahedra of sevens, the topological constraint Eq.~(\ref{constr_sphere}) 
becomes 
\be\label{ico_constr_sphere}
n_{+}^{\cal I}-n_{-}^{\cal I}=1 \ .
\ee
For a given configuration the energy is given by
\be\label{energy_ico_full}
E^{n{\cal I}}=C_{n{\cal I}}\frac{\pi K_0}{36} R^2 + 12(2n_{+}-1)\,E_{core}
\ee
where the $C$ coefficient is a function of 
$3(n_{+}^{\cal I}+n_{-}^{\cal I}-1)=6n_{+}^{\cal I}$ variables.
Let us first choose a  distinguished icosahedron with explicit coordinates 
\be\label{param_NP}
(\theta,\phi)\equiv\left\{(0,\mbox{any}),
(\gamma,\frac{2\pi k}{5})_{0\leq k \leq 4},
(\pi-\gamma,\frac{\pi}{5}+\frac{2\pi k}{5})_{0\leq k \leq 4},
(\pi,\mbox{any})\right\} \ ,
\ee
where $\gamma=\cos^{-1}(1/\sqrt{5})$. Each of the remaining icosahedra
may then be parametrized by the set of three Euler angles necessary to bring them
to the position described by Eq.~(\ref{param_NP}).

  \begin{figure}[htb]
  \epsfxsize=2 in \centerline{\epsfbox{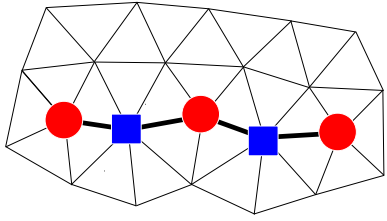}}
  \caption{Particle configurations near a finite grain boundary. Circles represent
  five-coordinated sites and squares represent seven-coordinated sites.}
  \label{fig__finite_grain}
  \end{figure}

The problem is to minimize over this set of Euler angles. We perform
this minimization using a direction set algorithm \cite{NumRec}. 
From the results shown in Table~\ref{tab__icos_different_l}, 
it is clear that the energy coefficient $C$ is reduced by 
the addition of defects. It is therefore favorable to form defects for
a sufficiently small core energy.

  \begin{table}[htb]
  \centerline{
  \begin{tabular}{|c|c||l|l|l||}
  \multicolumn{1}{c}{$n_{+}^{\cal I}$} & \multicolumn{1}{c}{Total} &
  \multicolumn{1}{c}{$C_{n{\cal I}_t}$} & \multicolumn{1}{c}{$a/R$}& 
  \multicolumn{1}{c}{$M$} \\\hline
   $1$  &  $12$  & $0.60$   & $\gamma$ & 12\\\hline 
   $2$  &  $36$  & $0.45$   & $0.09$   & 1791\\\hline
   $3$  &  $60$  & $0.38$   & $0.06$   & 4031\\\hline
   $4$  &  $84$  & $0.34$   & $0.03$   & 16124\\\hline
   $5$  & $108$  & $0.30$   & $0.02$   & 36279\\\hline
   $6$  & $132$  & $0.257$  & $0.02$   & 36279\\\hline
  \end{tabular} }
  \caption{Table of results for the minimum energy coefficient, as 
  defined in Eq.~(\ref{energy_ico_full}), obtained within the
  icosahedral approximation as a function of the number $n_{+}^{\cal I}$
  of icosahedral clusters of positive charge. The The penultimate column gives
  the average geodesic distance between neighboring charges. The last
  column gives the corresponding total number of particles, as
  estimated from Eq.(\ref{partno}).}
  \label{tab__icos_different_l}
  \end{table}

\begin{figure}[hp]
\centerline {\epsfxsize = 2.1in \epsfbox{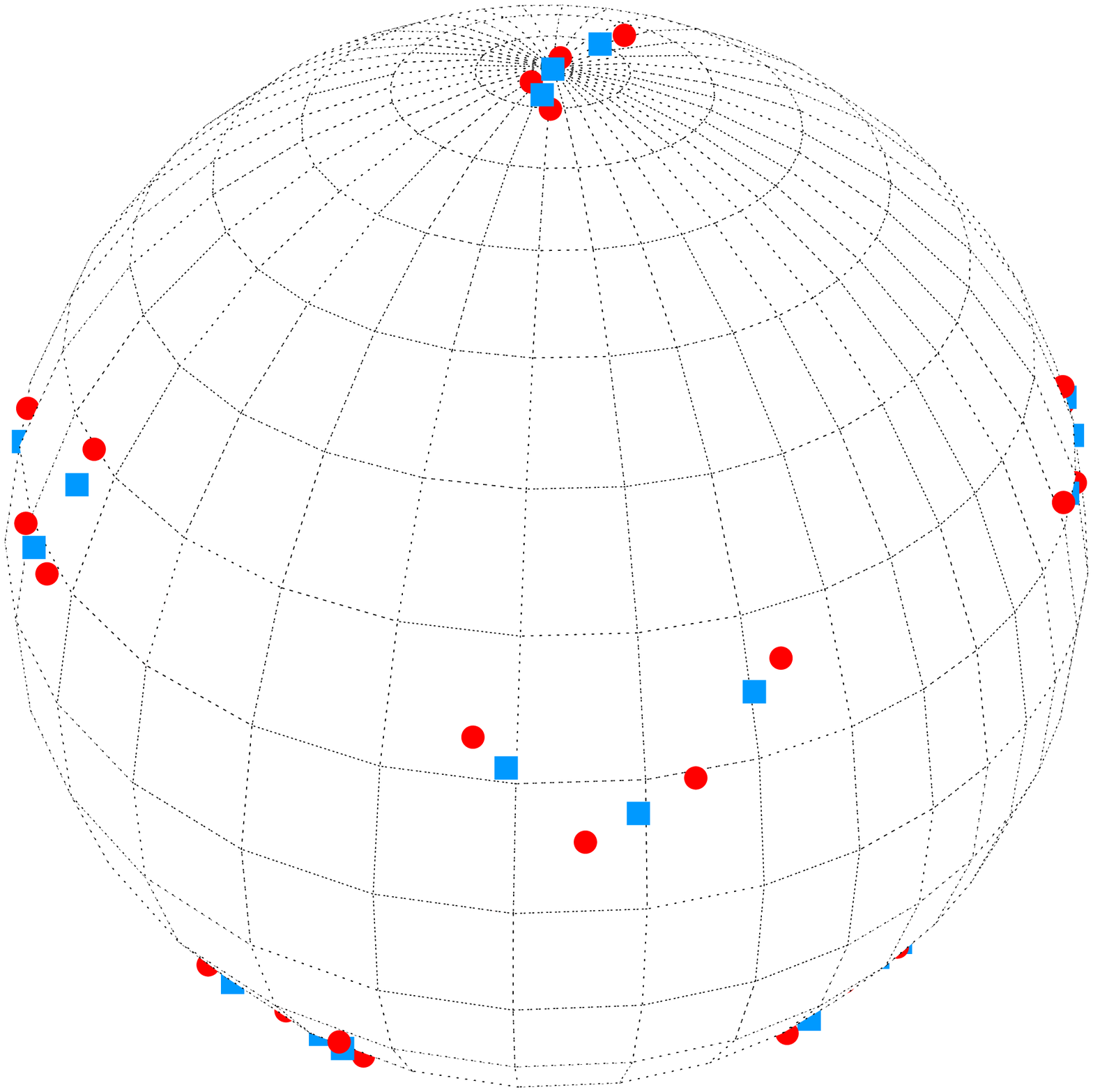}
\epsfxsize = 2.1in \epsfbox{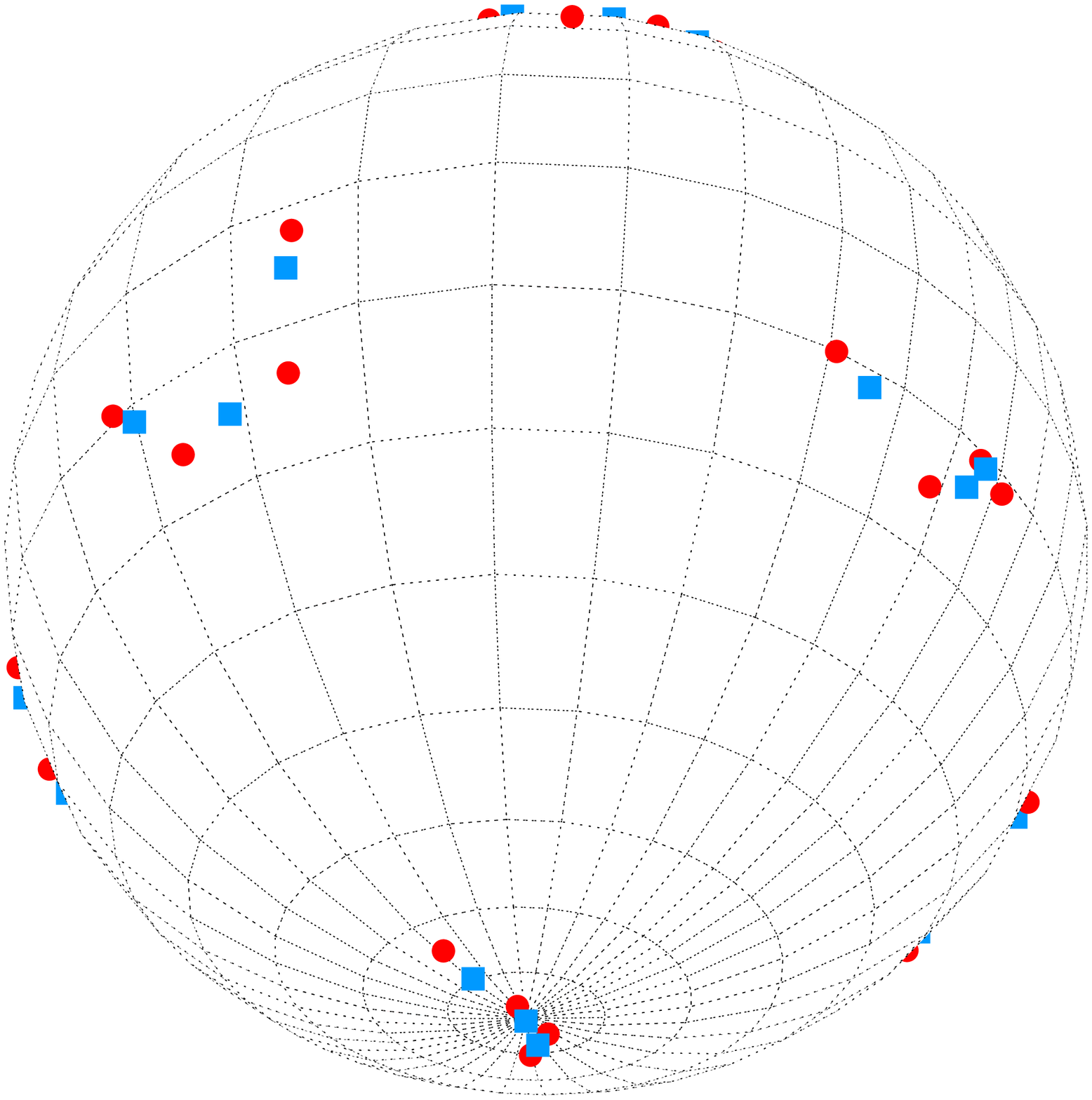 }}
\centerline {\epsfxsize = 2.1in \epsfbox{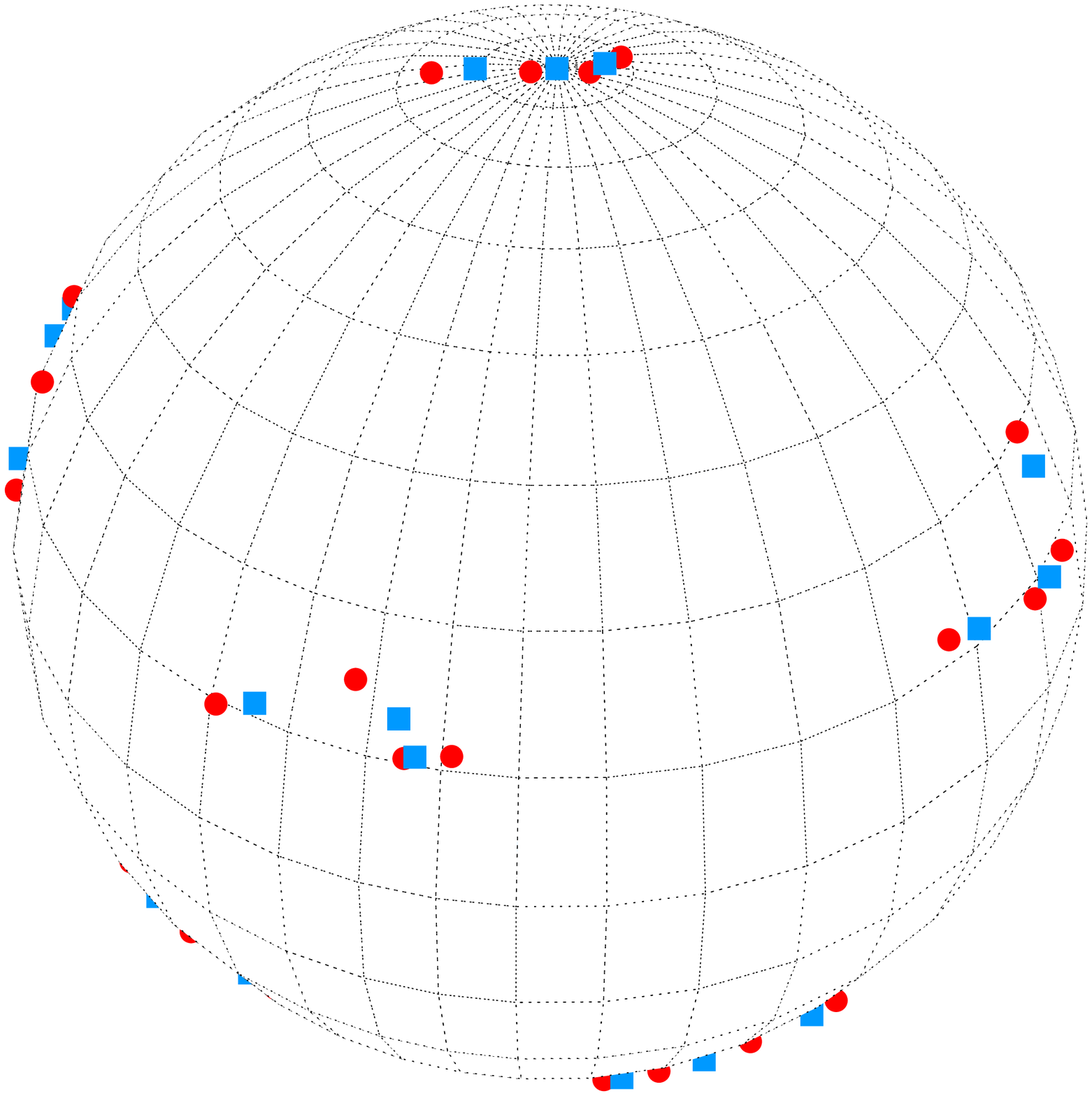}
\epsfxsize = 2.1in \epsfbox{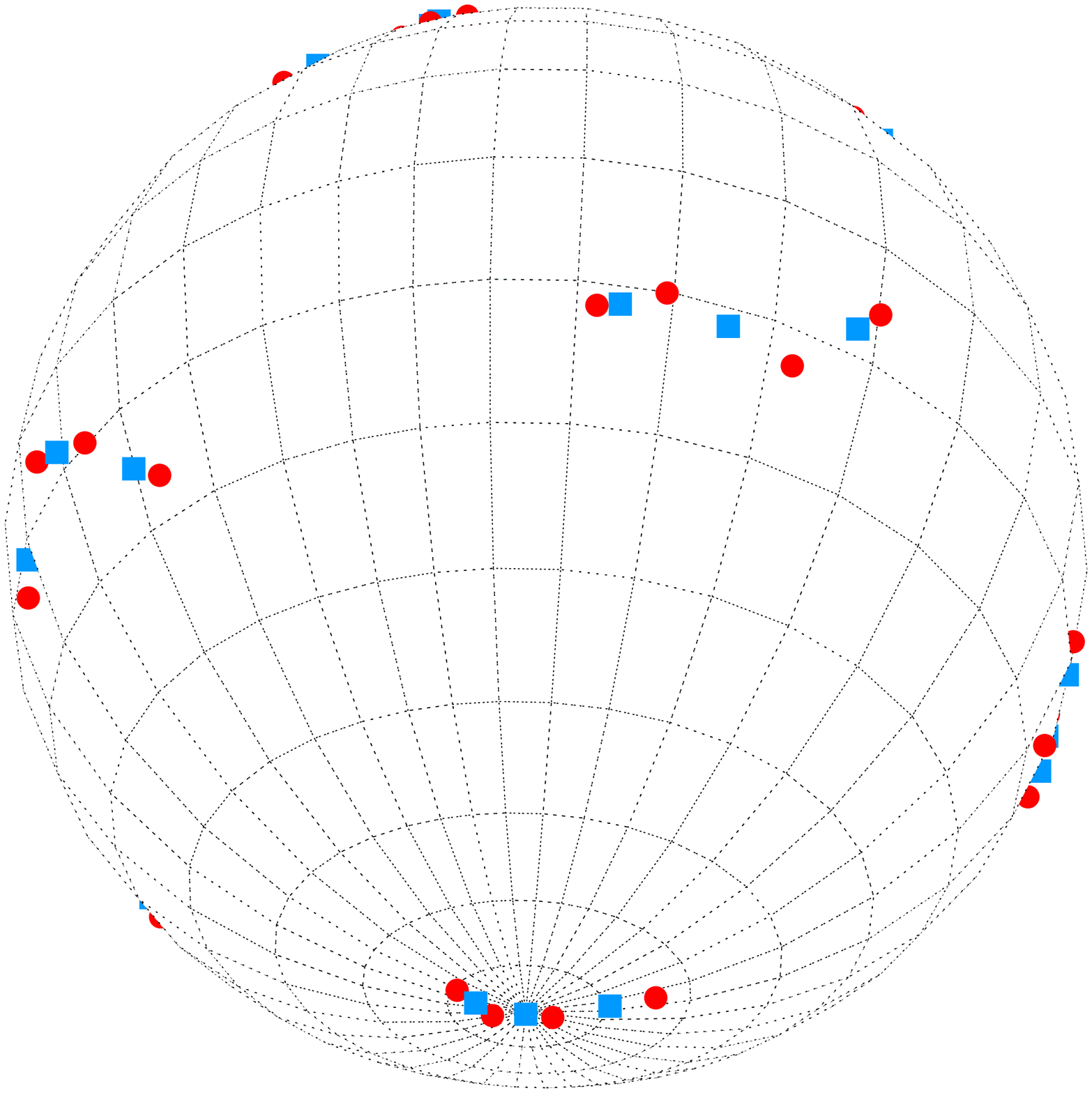}}
\centerline {\epsfxsize = 2.1in \epsfbox{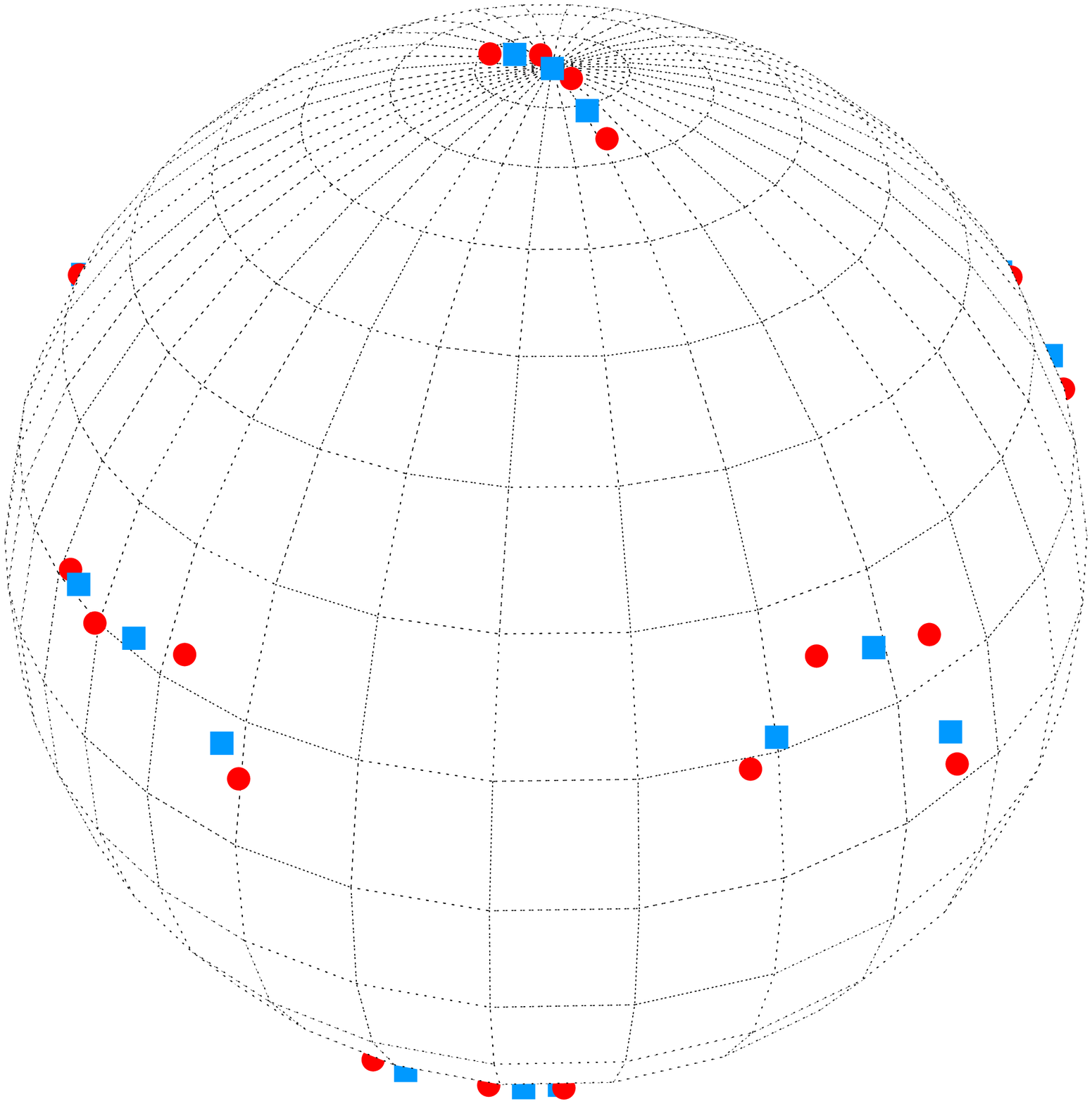}
\epsfxsize = 2.1in \epsfbox{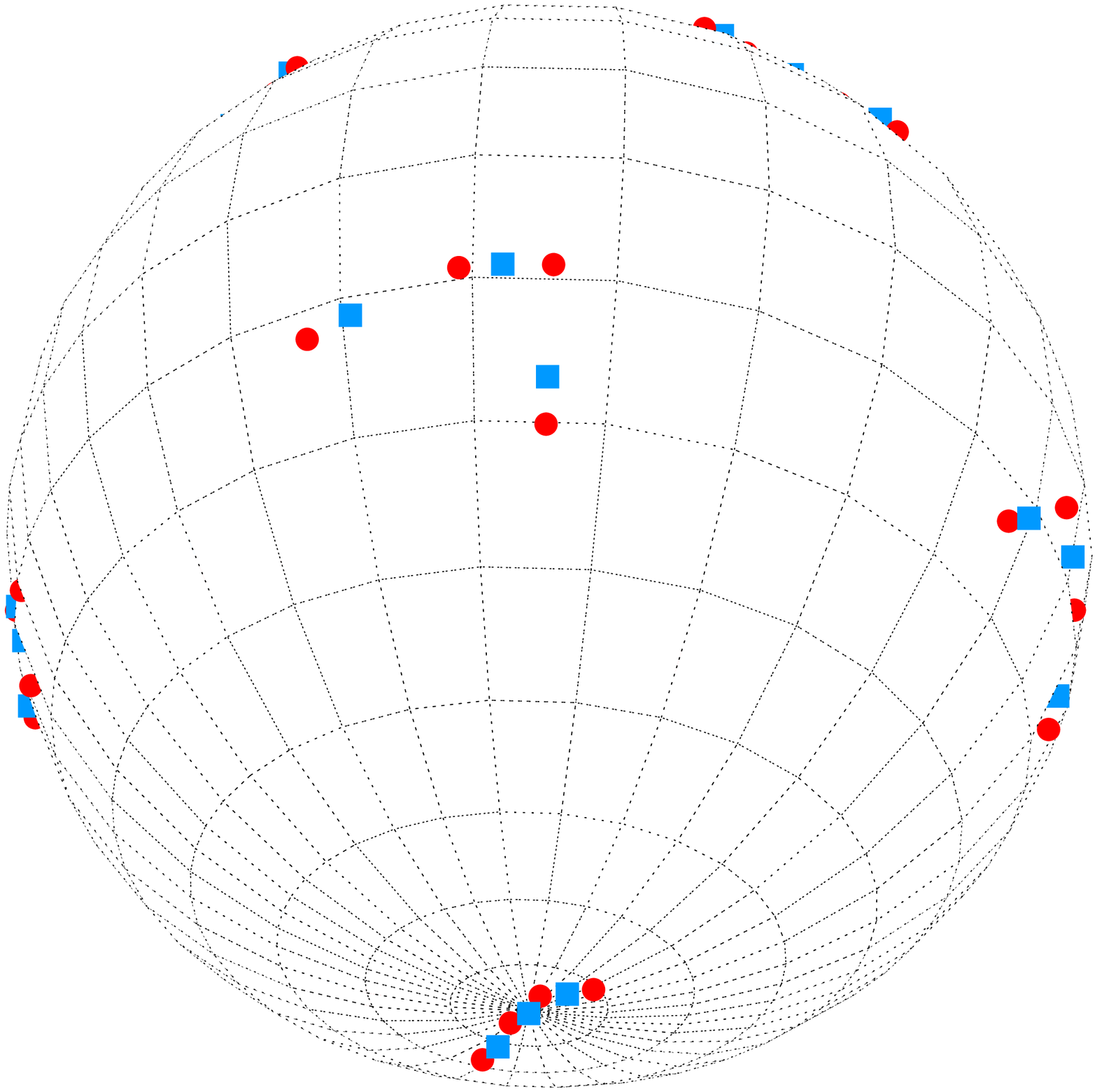}}
\caption{Six views of the ground state configuration for the
icosahedral solution with seven sets of icosahedral defect clusters.              
The top figure in each column shows the north and south pole
respectively. The subsequent views are obtained by successive rigid
body $120^{\circ}$ rotations of the entire sphere, using the
right-hand rule, about an axis running from the north pole to the south pole.}
\label{fig__ico7}
\end{figure}

Another important issue is the precise arrangement of defects in the
ground state. For $n_{+}<5$ we find the remarkable appearance of
finite grain boundaries {--} finite strings of interlaced fives and
sevens, as depicted schematically in Fig.~\ref{fig__finite_grain}.
These grain boundaries are not always perfectly linear,  although one
does find alternating disclination chains clustered along geodesic line
segments. Occasionally one finds dislocations, i.e. disclination
pairs, displaced from this geodesic by a few lattice spacings. 
The ground state we find, for the case $n_{+}=4$, is depicted
in Fig.~\ref{fig__ico7}, and nicely illustrates the above features.
Note that although the local structure of these grain boundaries
mimics that expected for flat space, the curvature of the sphere
allows these linear structures to terminate, consistent with the
discussion in the Introduction. 
 
For the case $n_{+}=5$ we still observe finite grain boundaries, but they
show a tendency to spiral. This tendency becomes 
more pronounced for the case $n_{+}=6$, where the finite 
strings evolve into more complicated structures.

The picture emerging then within the icosahedral approximation is that
small core energies favor a proliferation of defects. Below a critical number of
defects of order 100, the ground state is well described by twelve finite
grain boundaries, each one seeded by a defect in the original
icosahedron. Above the critical number of defects the finite grain
boundaries tend to branch and develop tentacles: the linear character of the
pattern is lost.

\subsection{String dominated regime}\label{sub_c3_sol}

\begin{figure}[hp]
\centerline {\epsfxsize = 2.25in \epsfbox{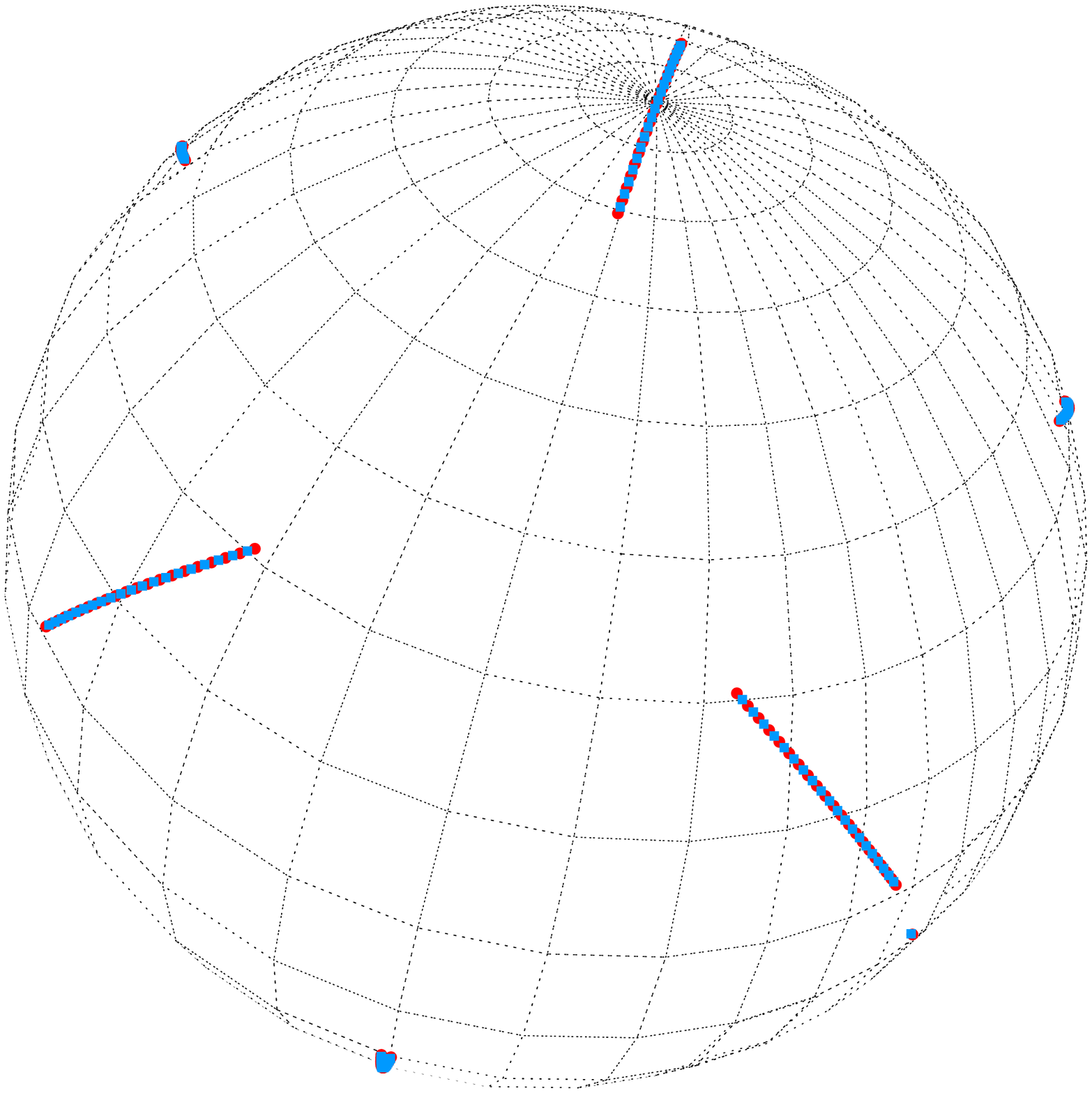}
\epsfxsize = 2.25 in \epsfbox{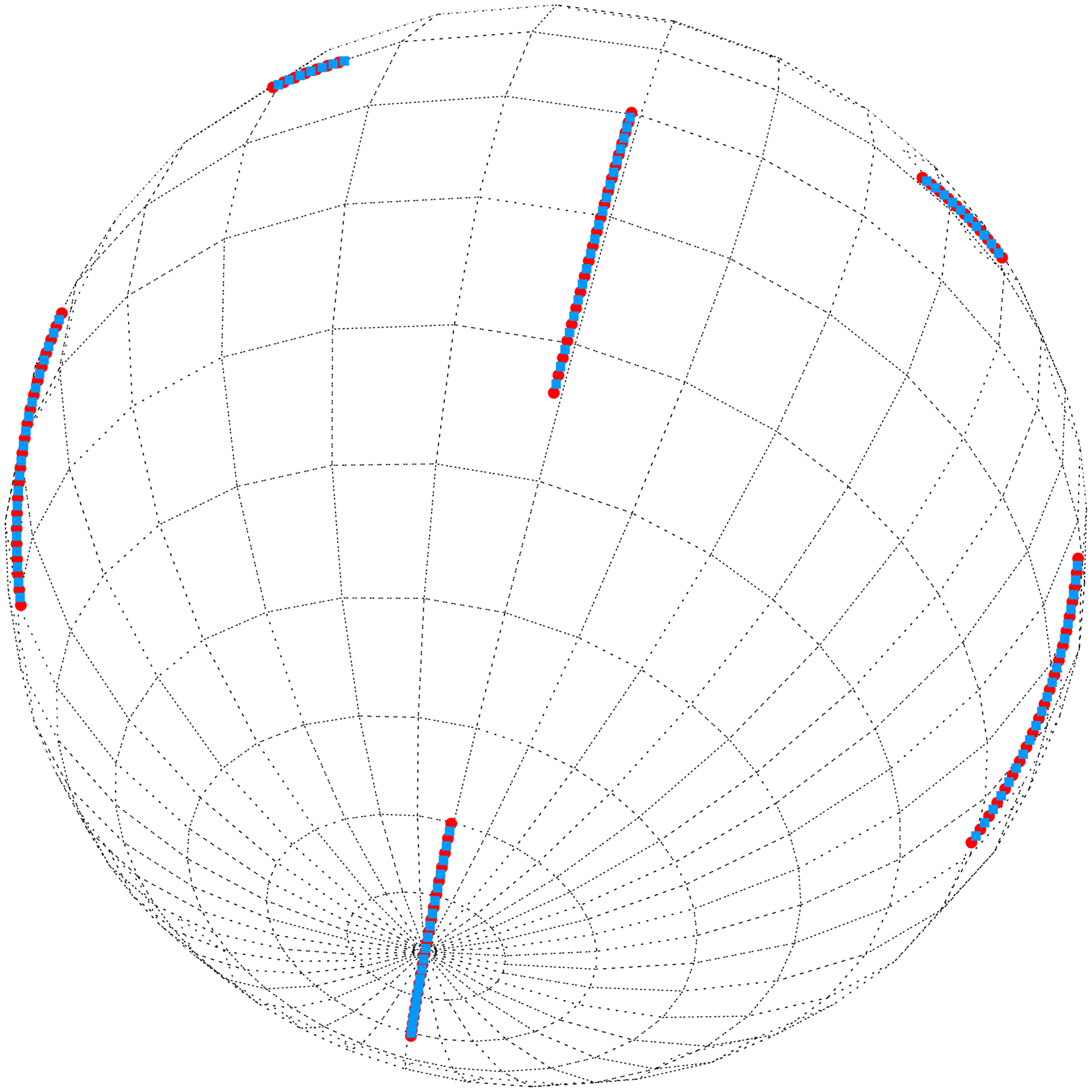 }} \centerline
{\epsfxsize = 2.25in \epsfbox{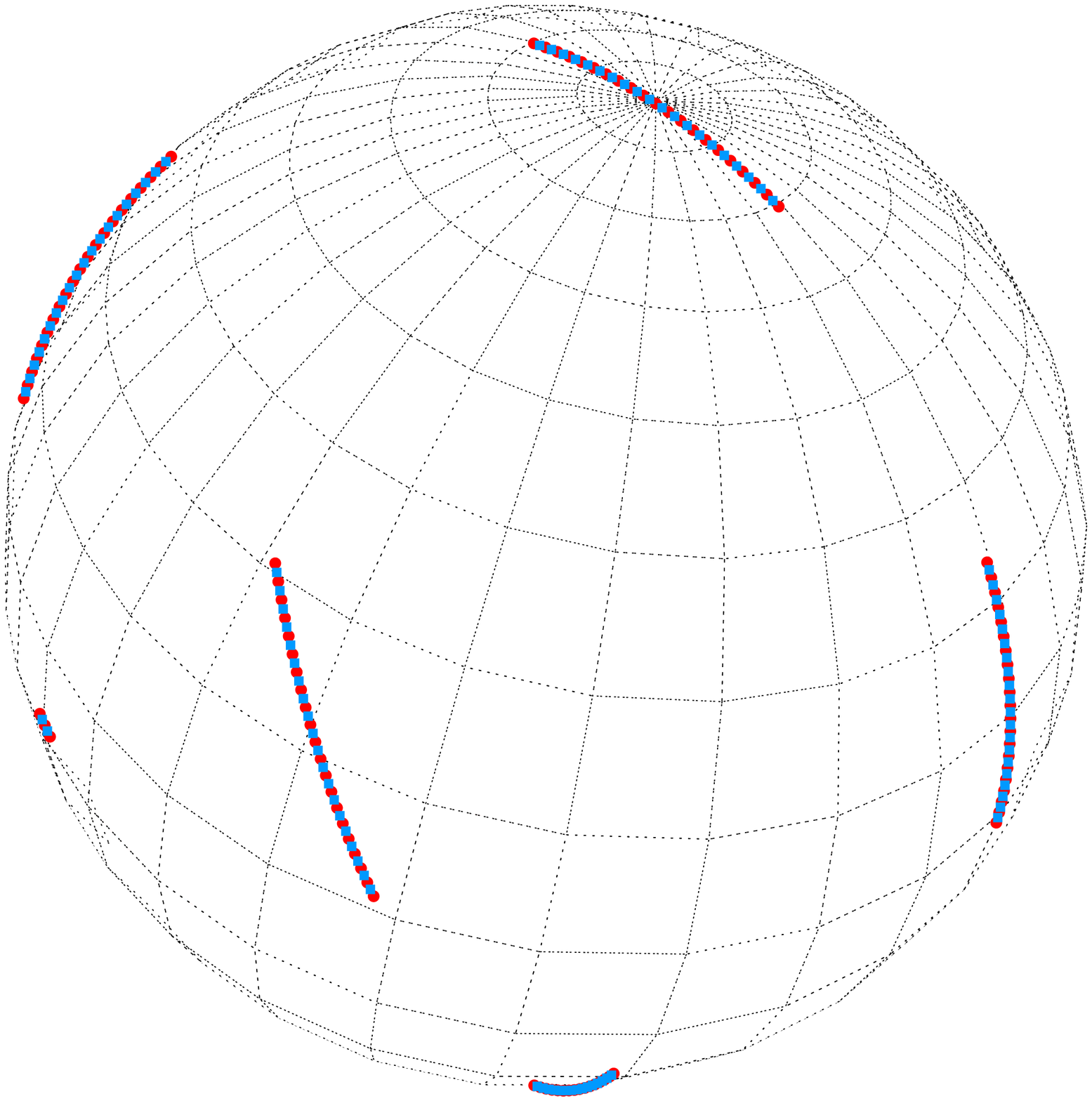} \epsfxsize = 2.25in 
\epsfbox{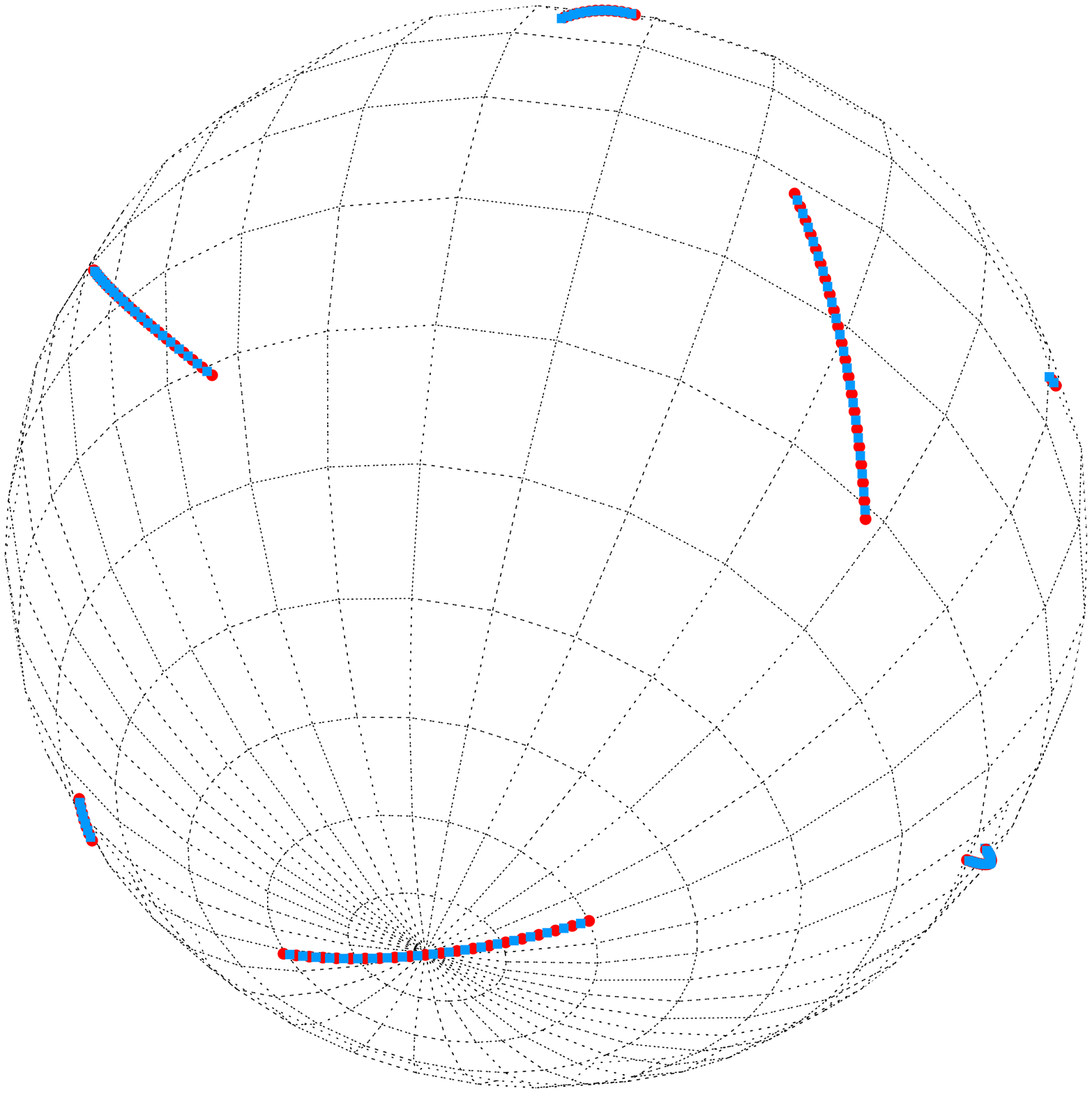}} \centerline {\epsfxsize = 2.25in
\epsfbox{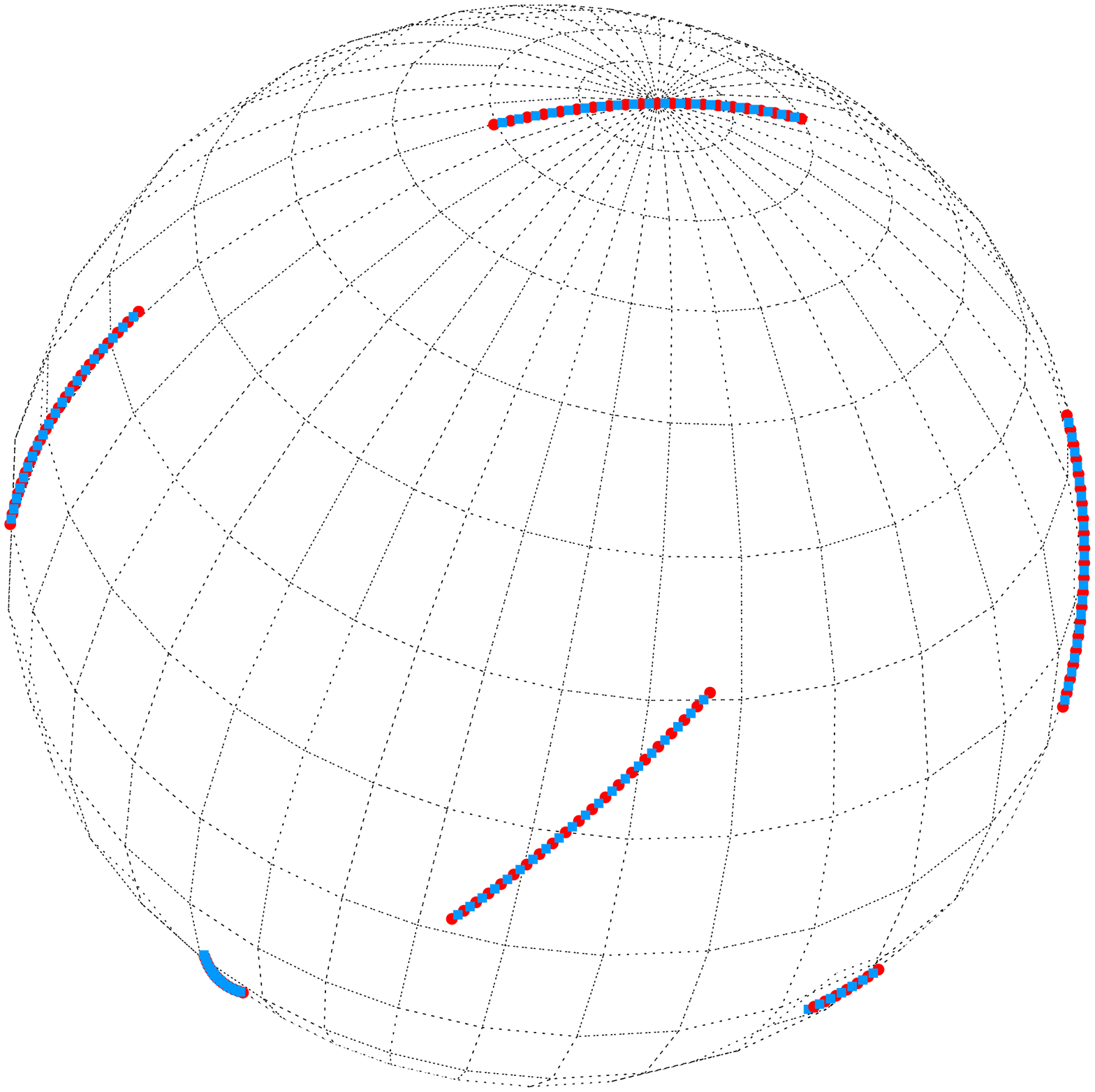} \epsfxsize = 2.25 in
\epsfbox{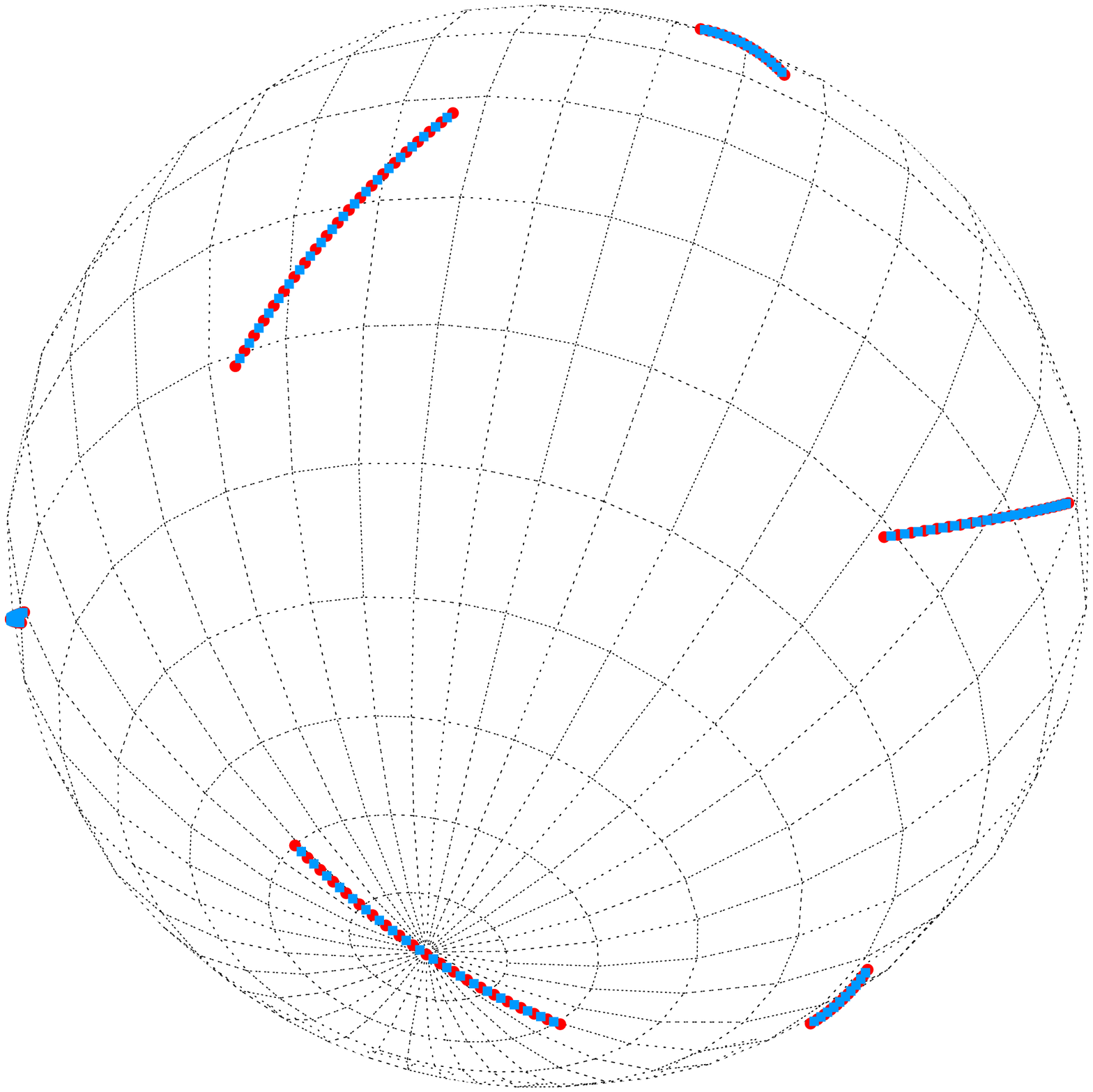}}
\caption{Six views of the ground state configuration for the $C_3$
solution with a large number of defects. The views are 
related as in Fig.~\ref{fig__ico7}.}
\label{fig__C3_final}
\end{figure}

In this section we examine the relative orientation of the finite
grain boundaries discussed in the previous section.  One ansatz is
provided by a solution having the form depicted in
Fig.~\ref{fig__C3_final}. There is an axis of three-fold rotational
($C_3$) symmetry at the center of the triangle formed by the geodesics
connecting the three nearest-neighbor disclinations of the icosahedral
array which forms the starting point of this variational ansatz. Finite
grain boundaries are constructed by adding defects along the geodesic
which joins the purely icosahedral sites with the center of
$C_3$-symmetry.  The midpoints of the grain boundaries form an
icosahedron.  The only free parameter in the model is the lattice 
spacing. This parameter may be fixed by minimizing the energy 
with respect to the lattice spacing, 
\be\label{cond_energy_zero} 
\frac{dE(a)}{d a}=0 \ .
\ee
The interpretation of this extremal lattice spacing is discussed in
Sec.~\ref{SECT__Ico}. 
\begin{table}[htb]
  \centerline{
  \begin{tabular}{|c||l|l||}
  \multicolumn{1}{c}{Total} &
  \multicolumn{1}{c}{$C$} & \multicolumn{1}{c}{$a$} \\\hline
       $12$  & $0.60$   & $\gamma$ \\\hline 
       $36$  & $0.44$   & $0.121$  \\\hline
       $60$  & $0.37$   & $0.085$  \\\hline
       $84$  & $0.34$   & $0.062$  \\\hline
      $108$  & $0.32$   & $0.051$  \\\hline
      $132$  & $0.31$   & $0.042$   \\\hline
      $252$  & $0.28$   & $0.024$   \\\hline
      $492$  & $0.26$   & $0.012$   \\\hline
      $972$  & $0.255$  & $0.006$   \\\hline
  \end{tabular} }
  \caption{The minimum energy coefficient (see
  Eq.~(\ref{energy_ico_full})) for the $C_3$ solution, 
  as a function of the total number of defects. 
  The last column gives the lattice spacing $a$ as determined from 
  Eq.~(\ref{cond_energy_zero}).}
  \label{tab__string_different_l}
  \end{table}

\begin{figure}[hp]
\centerline {\epsfxsize = 2.25in \epsfbox{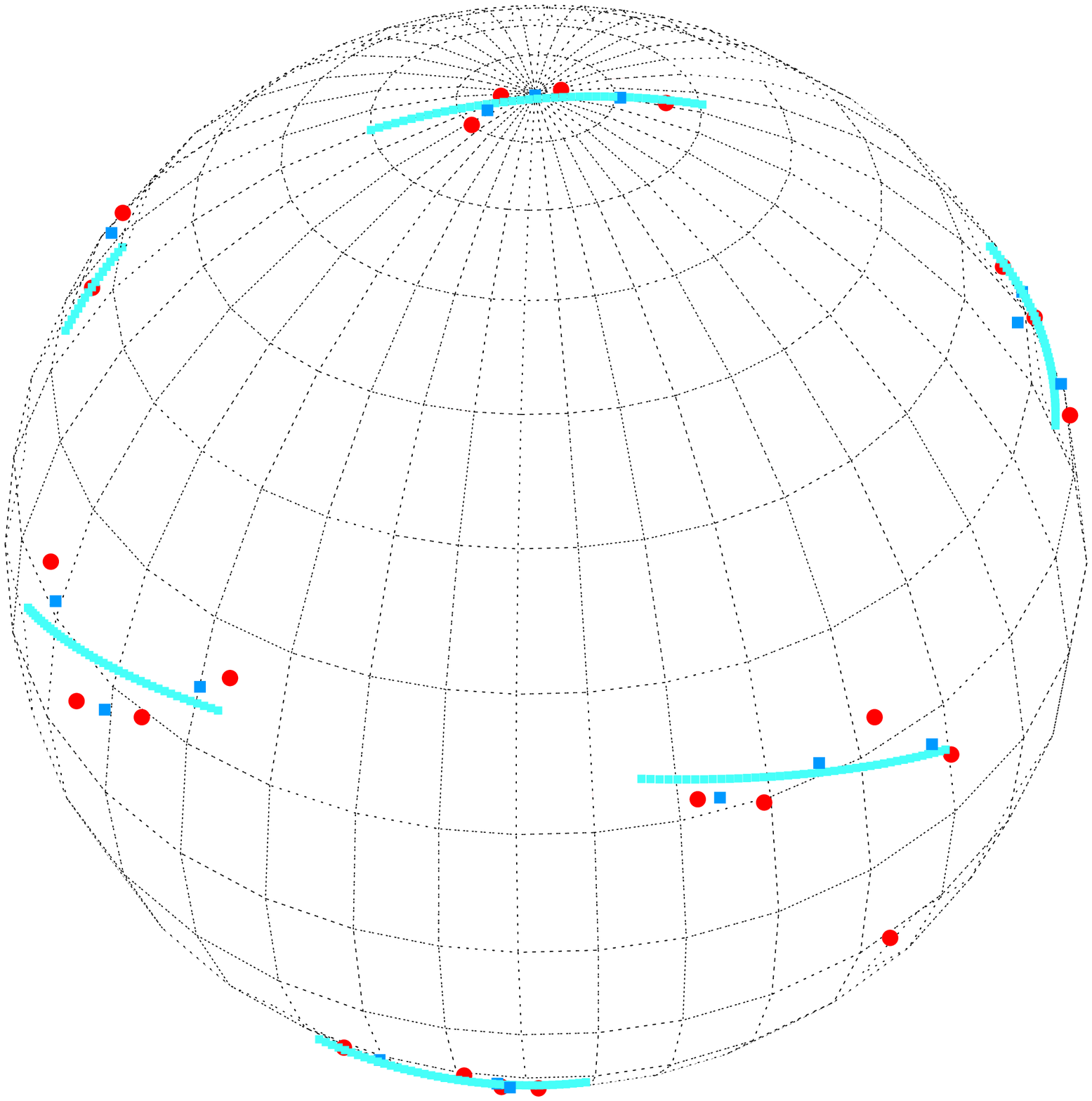}
\epsfxsize = 2.25 in \epsfbox{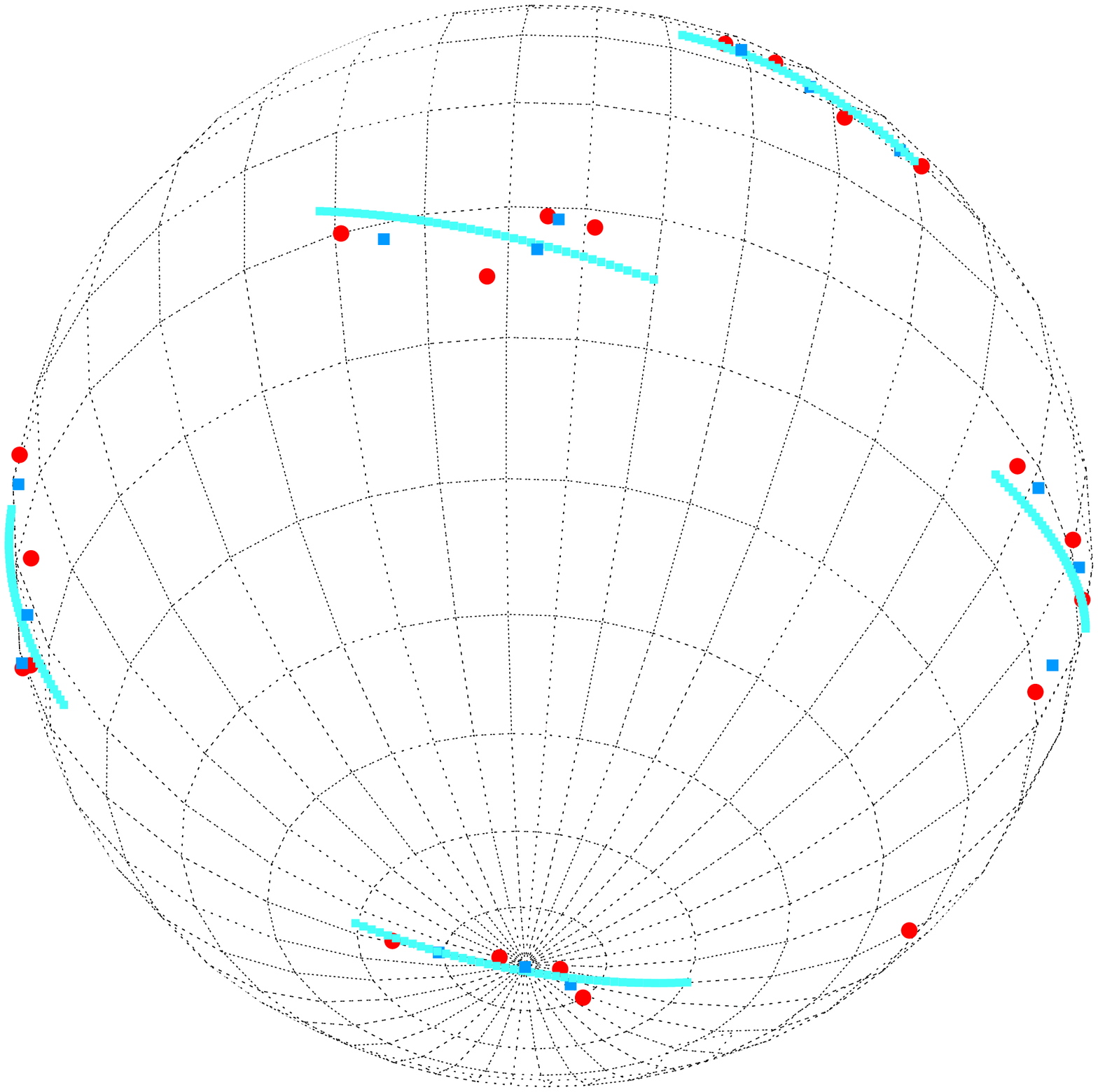 }}
\centerline {\epsfxsize = 2.25in \epsfbox{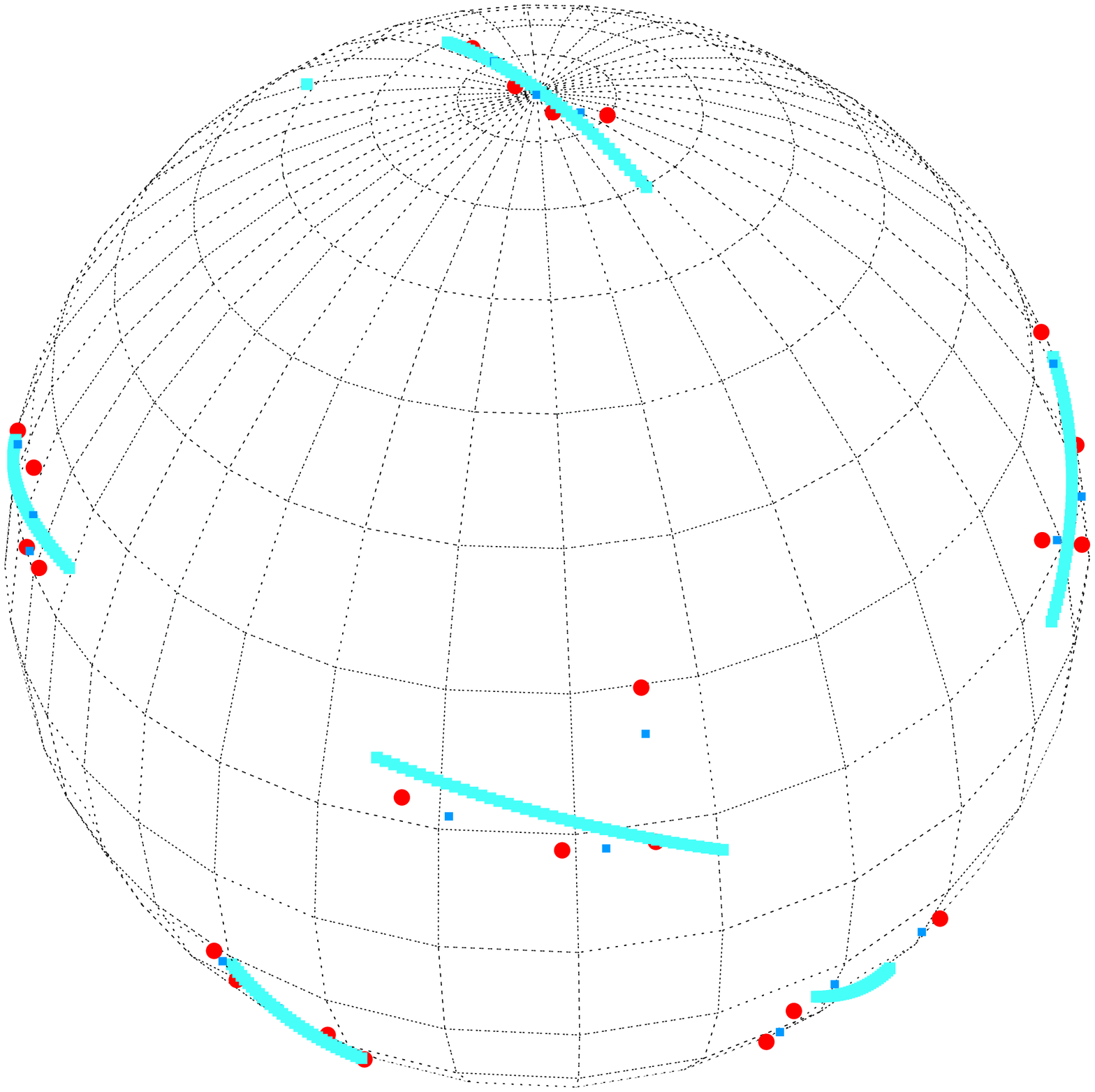}
\epsfxsize = 2.25 in \epsfbox{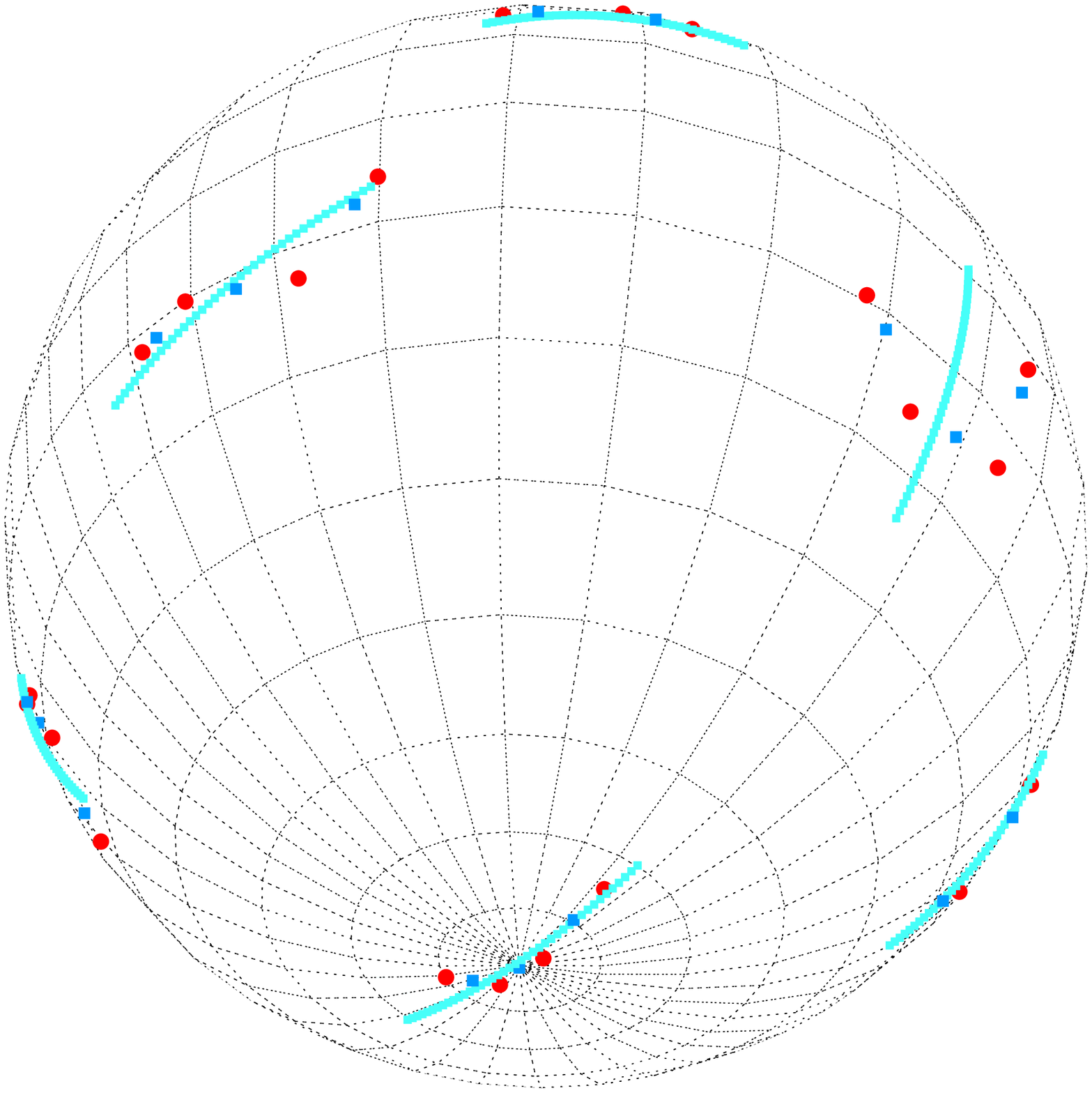}}
\centerline {\epsfxsize = 2.25in \epsfbox{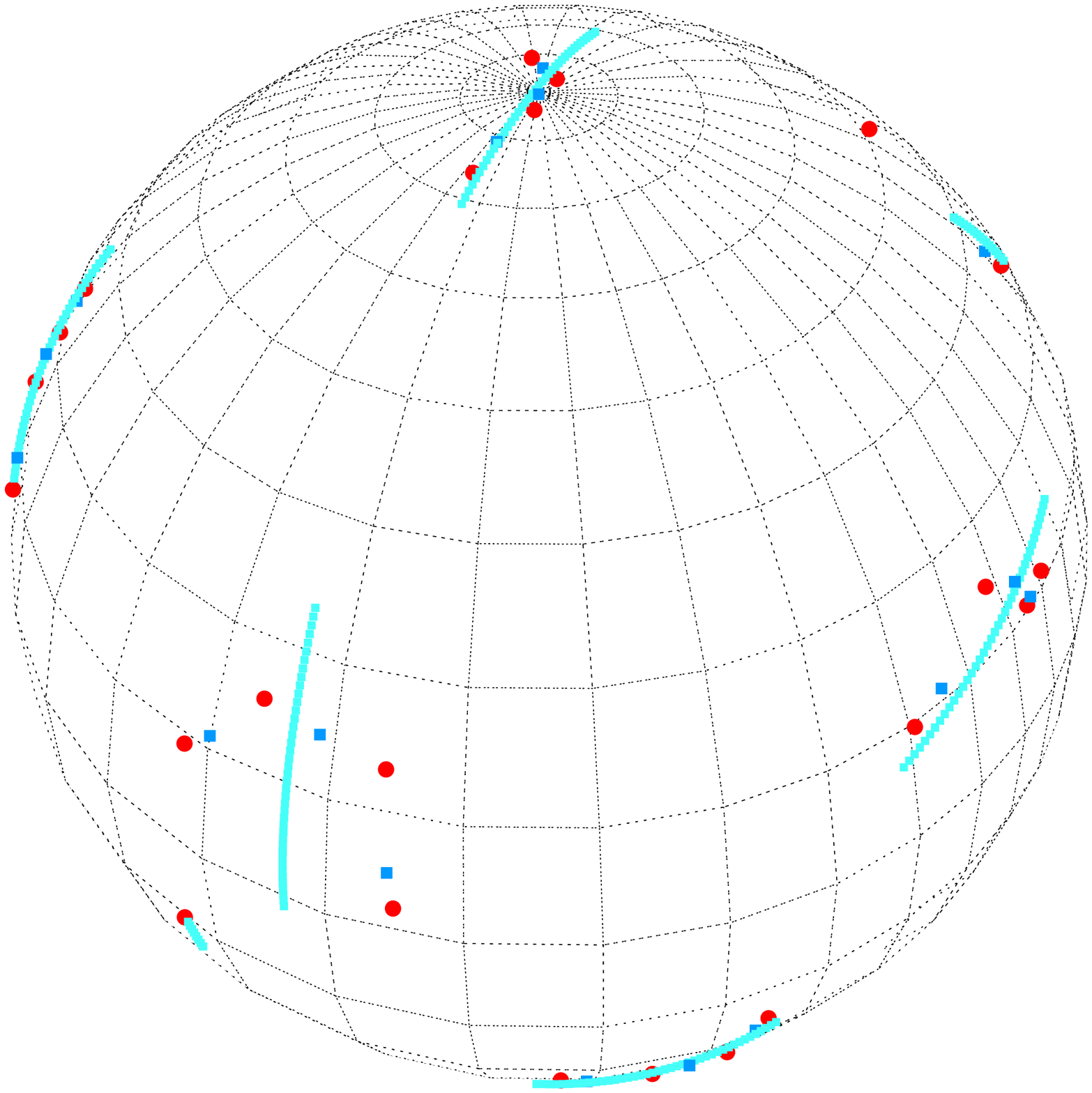}
\epsfxsize = 2.25 in \epsfbox{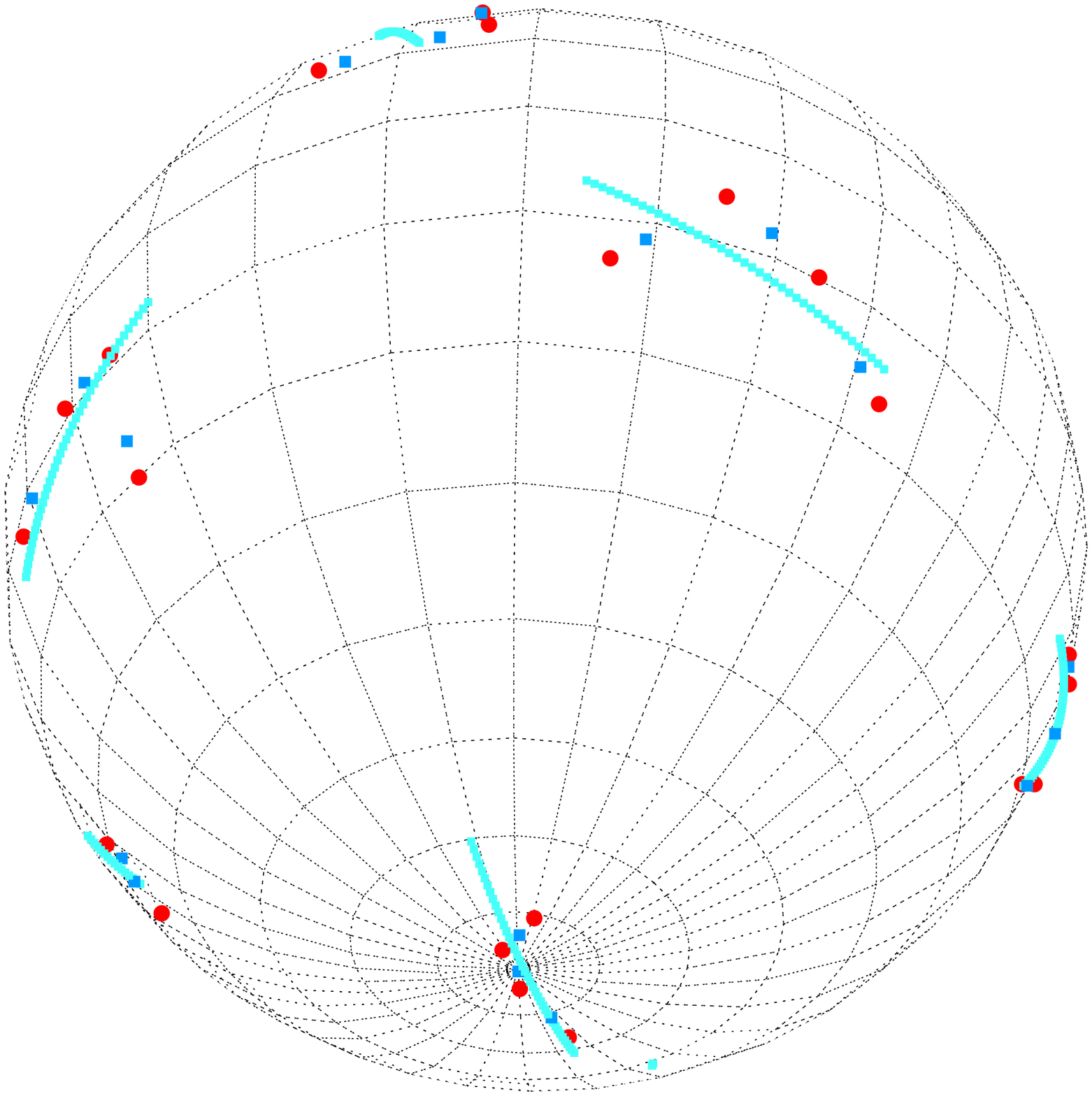}}
\caption{Six views of the ground state configuration for the
superposition of the $C_3$ solution and the icosahedral solution Fig.~\ref{fig__ico7}.
The views are related as in Fig.~\ref{fig__ico7}.}
\label{fig__C3_ico7}
\end{figure}

  The results from this minimization are shown in
  Table~\ref{tab__string_different_l}. When the total number of
  defects is less than a critical value (approximately 110) this $C_3$
  solution has energies slightly lower than those found within the
  icosahedral approximation. This is remarkable if one recalls that
  this $C_3$-symmetric solution is obtained by minimizing with respect
  to only {\em one} parameter, the lattice spacing.  The results
  obtained from the icosahedral approximation itself are in rough
  agreement with this very simple ansatz, as apparent from
  Fig.~\ref{fig__C3_ico7}.  Table~\ref{tab__string_different_l} also
  makes clear that there is little gain in energy when the total
  number of defects exceeds the critical value, even in the limit of a
  very large number of defects. This is consistent with the picture
  that purely linear finite grain boundaries are replaced by more
  complicated structures when the number of defects is large.
  
A more sophisticated treatment of {\em strings}, motivated by the
discussion of dislocations in the Introduction, would build the grain
boundaries from disclination dipoles with fixed size and then allow a
variable spacing between these dislocations. We hope to pursue this
approach in a future publication.

\subsection{Large number of defects}

\begin{figure}[ht] \epsfxsize=1 in
  \centerline{\epsfbox{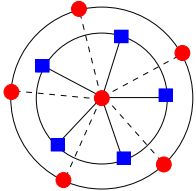}} \caption{Example of a simple
  star defect with two rings, one of fives (circles) and one of sevens
  (squares).}  
  \label{fig__stars} 
  \end{figure}

  We now have conclusive evidence that additional defects can lower
  the total energy of the system for small core energies.  Defects
  will then proliferate and form highly complicated
  patterns.  A detailed investigation of this regime is in
  progress, with complete results to be presented elsewhere. In this
  section we present one example of a branched structure that has
  lower energy than any of the linear structures considered so far.
  The structure we analyze consists of defects arranged in star
  patterns, or {\em pentagonal buttons} in the terminology of Toomre\cite{ALAR}
  (see Fig.~\ref{fig__stars}). To study these structures we construct
  rings of five-disclinations forming a pentagon with its center at the
  position of the icosahedron, as shown in Fig.~\ref{fig__stars}.  As
  free parameters we leave the radius of each ring, and the angle each
  ring forms with a given geodesic joining the center of a star to
  that of its neighbor. The topological constraint
  Eq.~(\ref{constr_sphere}) implies that there must be the same number
  of rings of fives as sevens.

\begin{figure}[hp]
\centerline {\epsfxsize = 2.25in \epsfbox{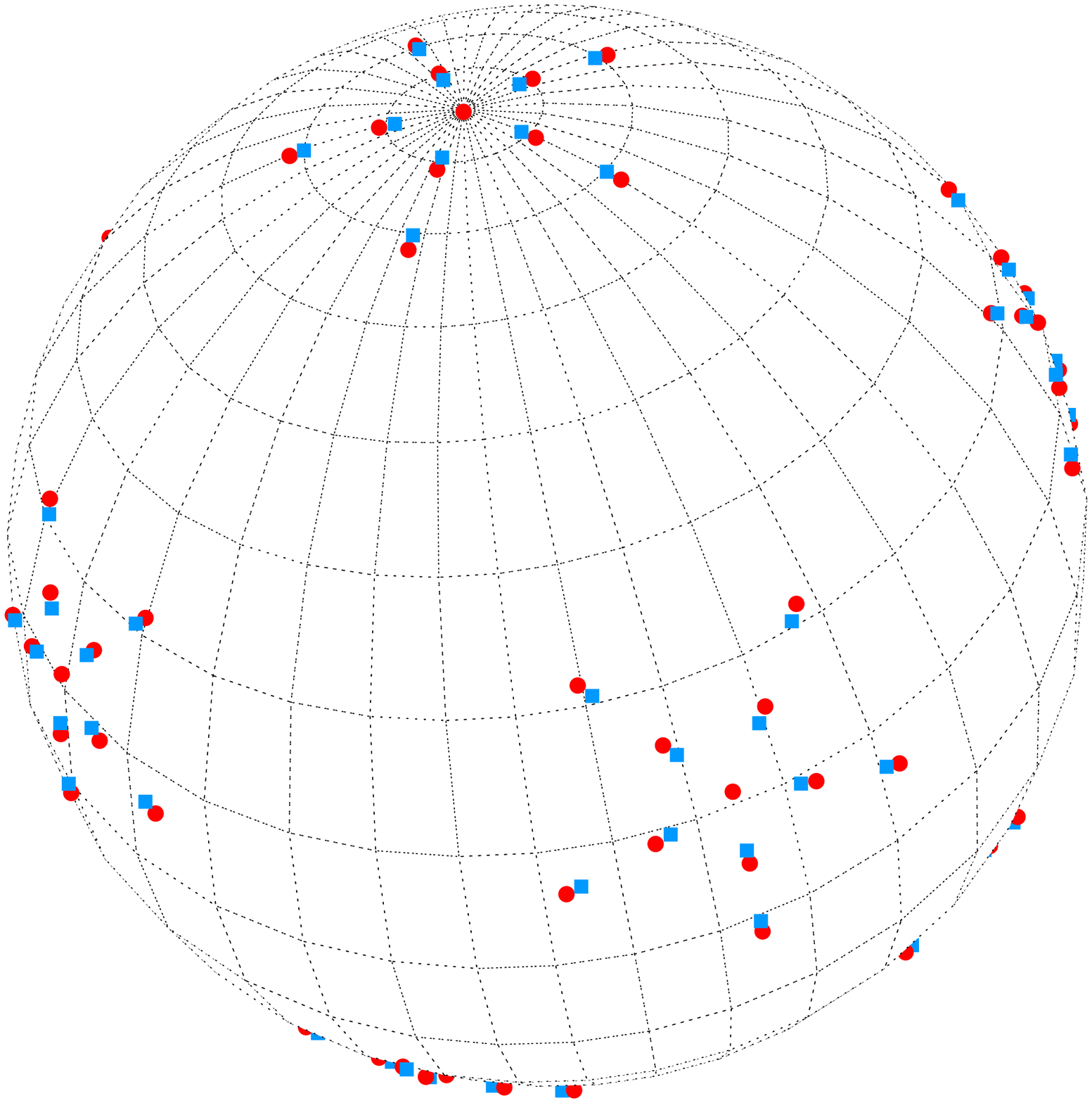}
\epsfxsize = 2.25 in \epsfbox{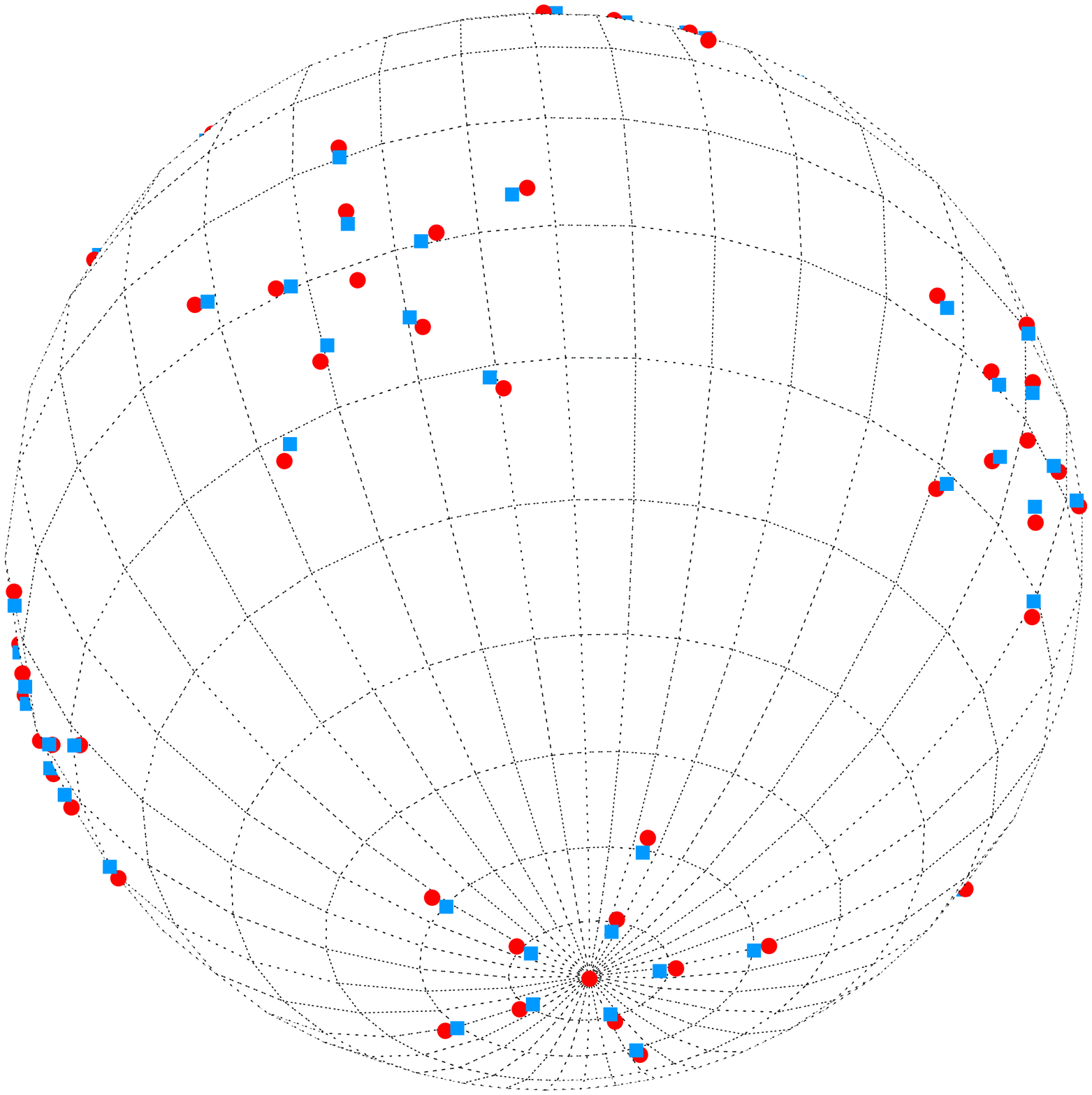 }}
\centerline {\epsfxsize = 2.25in \epsfbox{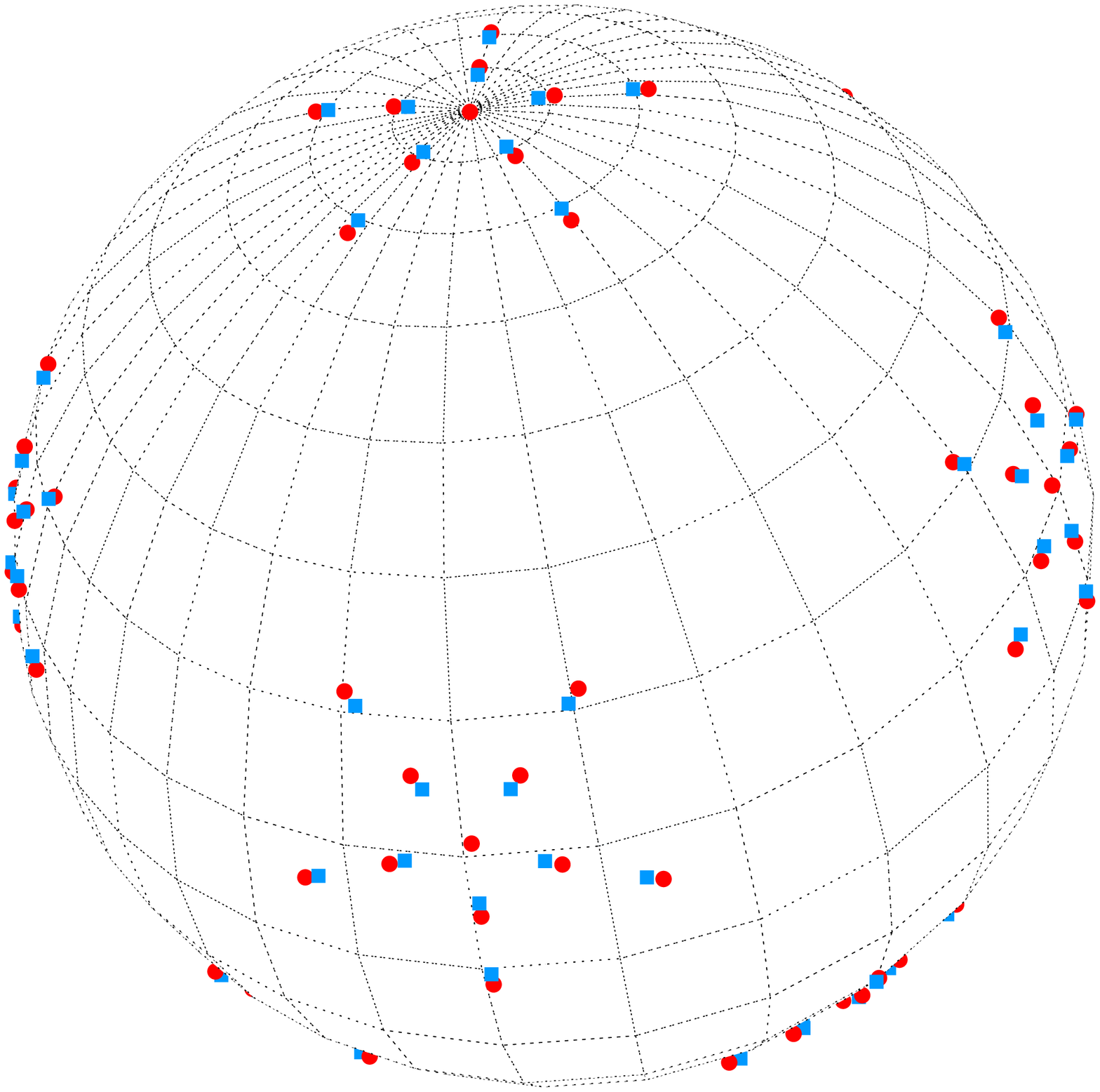}
\epsfxsize = 2.25 in \epsfbox{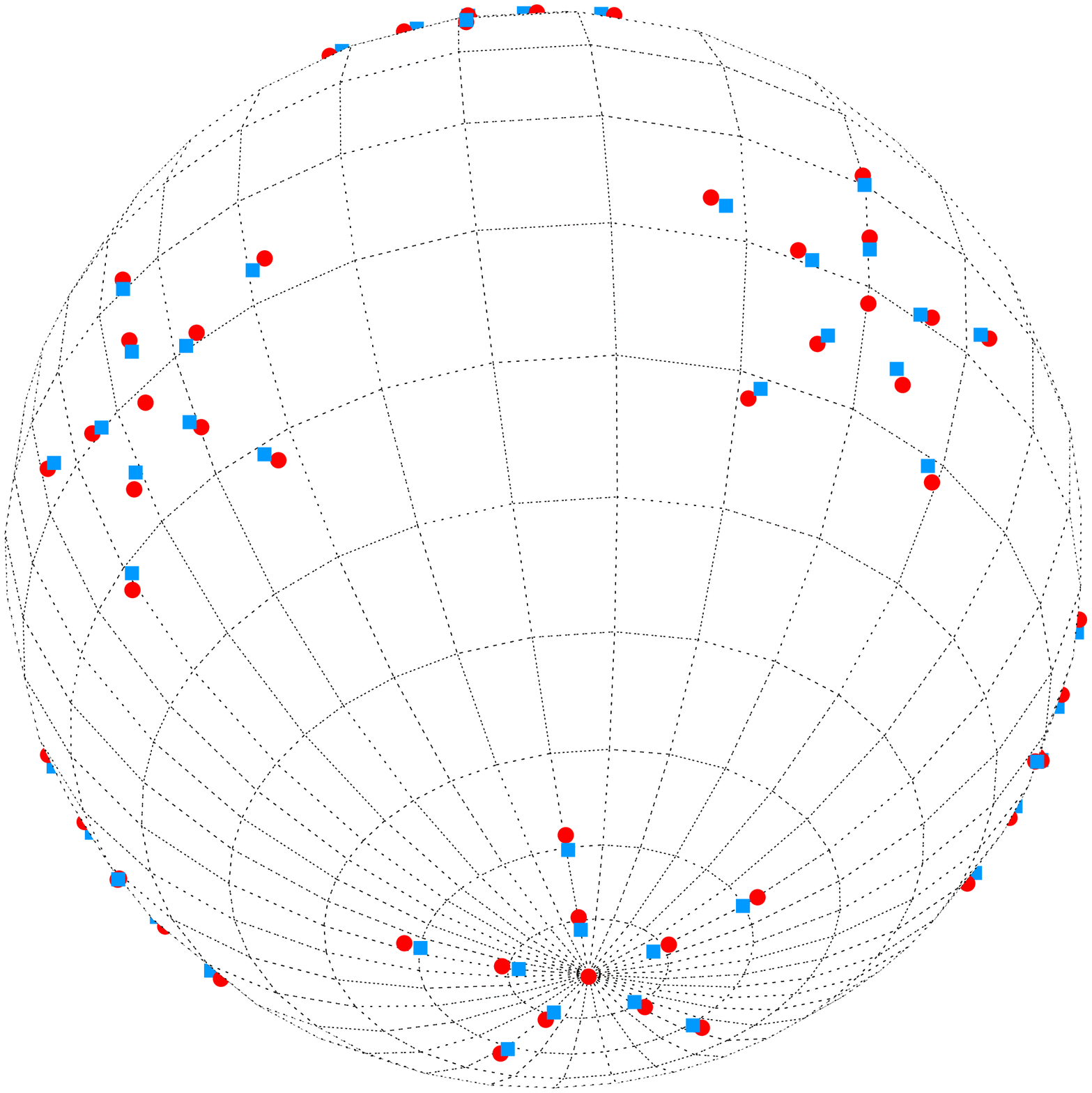}}
\centerline {\epsfxsize = 2.25in \epsfbox{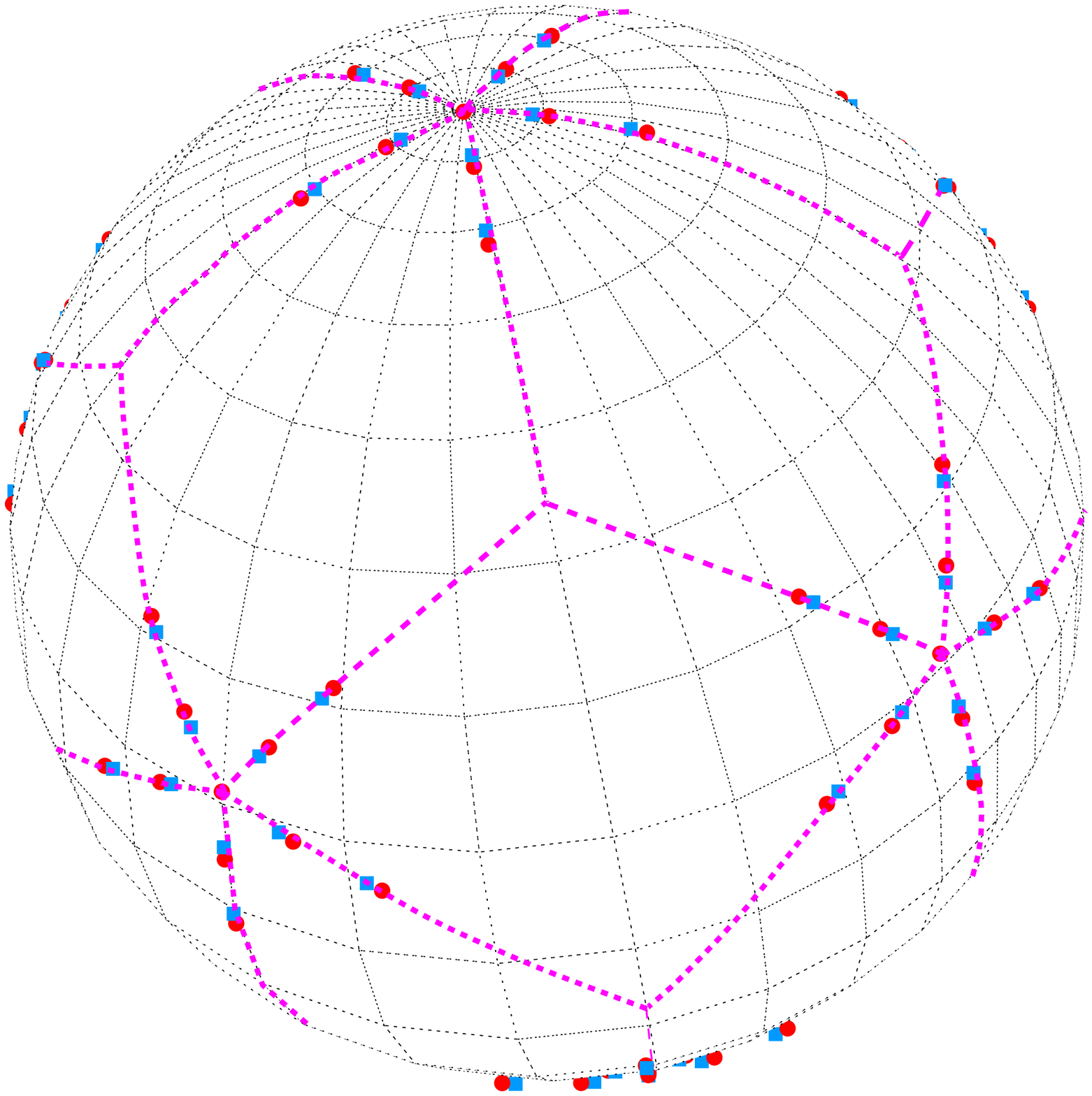}
\epsfxsize = 2.25 in \epsfbox{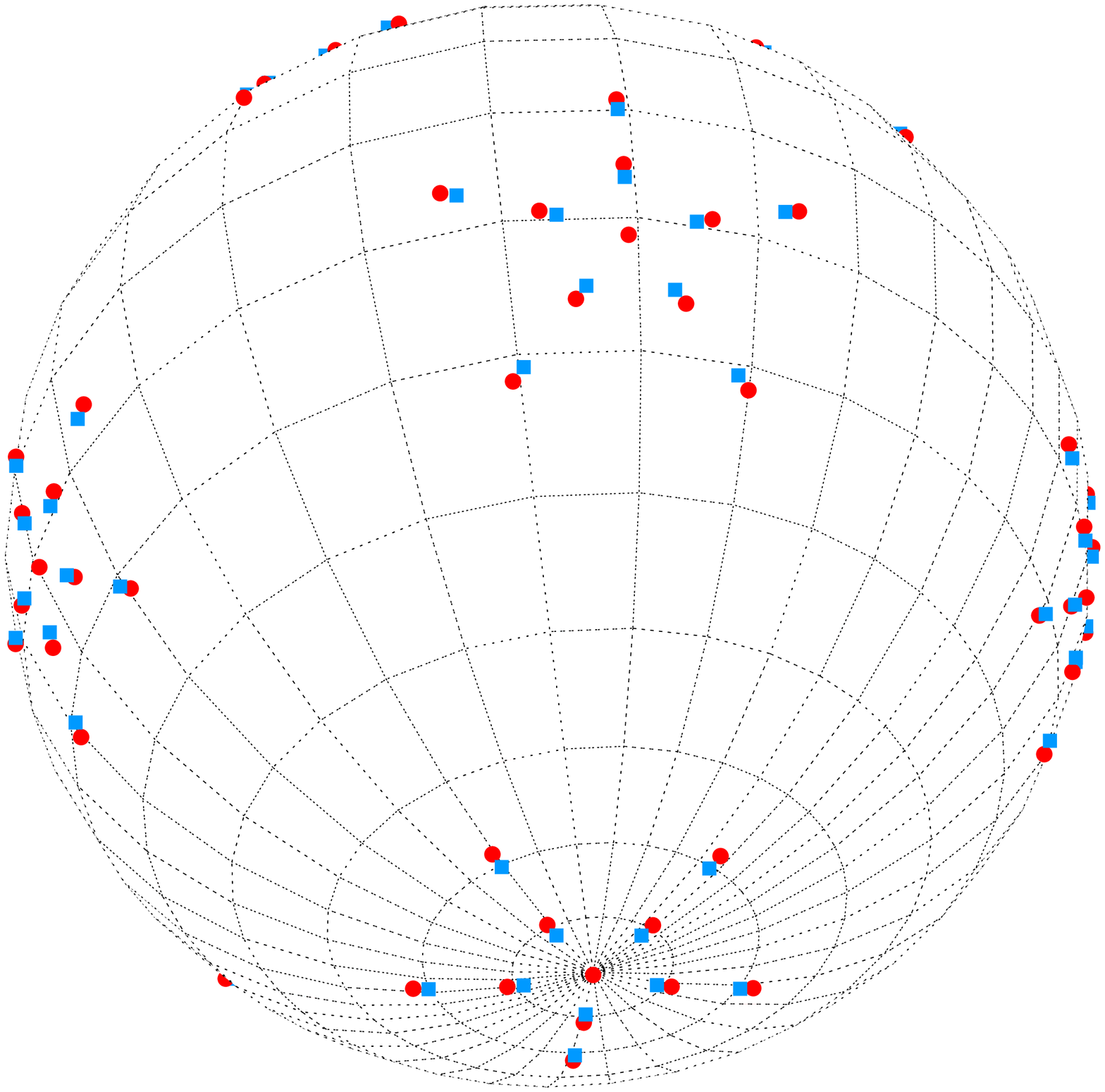}}
\caption{Six views of the ground state configuration for four rings of
{\em pentagonal buttons}. The views are related as in
Fig.~\ref{fig__ico7}. The bottom left view shows the associated
rhombic tiling (the rhombic tricontahedron) of the sphere.}
\label{fig__pent4}
\end{figure}

  From Table~\ref{tab__pent_different_l} we see the energy for $132$
  defects is marginally lower than the corresponding value for the
  icosahedral approximation. For more defects the star clusters have
  significantly lower energy than the $C_3$ solution. It is 
  remarkable that all the disclinations in this ground state solution,
  other than the twelve seed disclinations, bind to form {\em radial}
  dislocations as illustrated in Fig.~\ref{fig__pent4}). Furthermore, 
  the relative orientation of the different rings conform to a rhombic
  tiling of the sphere consisting of $30$ completely regular diamonds
  (the {\em rhombic tricontahedron}), 
  as shown in the bottom left picture of Fig.~\ref{fig__pent4}. 
  Note that by minimizing the defect elastic energy we obtain a
  {\em dynamically generated} particle spacing, for a fixed sphere
  radius, which optimizes the given structure.  
  Further investigation of these pentagonal buttons, as well as 
  other more involved structures, will be presented in the future.
  
  \begin{table}[htb]
  \centerline{
  \begin{tabular}{|c||l|l||}
  \multicolumn{1}{c}{Total} &
  \multicolumn{1}{c}{$C$} & \multicolumn{1}{c}{$a/R$} \\\hline
      $132$  & $0.255$  & $0.04$   \\\hline
      $252$  & $0.170$  & $0.025$   \\\hline
  \end{tabular} }
  \caption{ Table of results for the minimum energy coefficient, as 
  defined in Eq.~(\ref{energy_ico_full}) obtained within star defects 
  as a function of the total number of defects. The last column gives the 
  value of the particle spacing $a$.}
  \label{tab__pent_different_l}
  \end{table}

  It is natural, at this point, to ponder the nature of the ground state in
  the limit of an infinite number of defects with vanishing core
  energy or, equivalently, in the limit $R/a \rightarrow \infty$. 
  We have, in fact, already addressed this question in
  subsect.~\ref{sub_sect_spher}, where we proved that the only zero
  energy solution is an arrangement of defects
  $\left\{q_i,(\theta_i,\phi_i)\right\}_{i=1,\cdots}$ satisfying
  Eq.~(\ref{zero_screen}). In this case the defect density would be
  fully rotational invariant and screen out the Gaussian curvature
  completely. Currently we find solutions that seem to be
  converging to this limiting case, but it is open as to how accurately
  one can achieve the desired limit $C=0$.

\section{Instabilities of Icosahedral Lattices}
\label{SECT__Ico}
  
Our discussion so far has focused entirely on analyzing the
distribution of topological defects on the sphere. We turn our
attention now to the implications for the underlying lattice
structure. We thus take into account the regular six-fold coordinated
nodes as well as the defects and examine the resultant lattices.

\begin{figure}[htb]
  \epsfxsize=3 in \centerline{\epsfbox{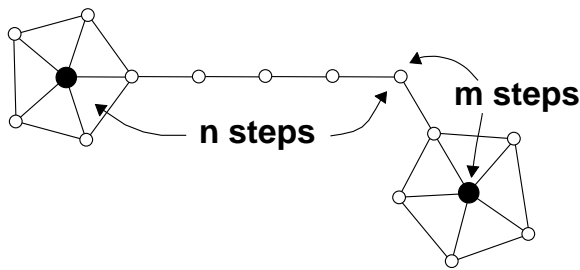}}
  \caption{The construction of a type (n,m) icosadeltahedral
  lattice. The filled circles indicate two nearest-neighbor 
  five-fold disclinations.} 
  \label{fig__nmicosa}
  \end{figure}

In the limit of large core energies our model predicts twelve
disclinations forming an icosahedron. Lattices whose only defects are
twelve positive disclinations sitting at the vertices of an icosahedron
may be constructed easily, since they are characterized by the path
between two nearest-neighbor disclinations. For a type $(n,m)$ lattice
this path consists of $n$ straight steps from a given disclination, a
$120^{\circ}$ turn, and then $m$ more straight steps to the
nearest-neighbor disclination (see Fig.~\ref{fig__nmicosa}). 
The total number of particles $M$ within this
$(n,m)$ (icosadeltahedral) lattice  is
\be\label{num_atoms_nm}
M=10(m^2+n^2+mn)+2 \ .
\ee
Within our model, the energy for these configurations has been computed 
in Eq.~(\ref{energy_icos_ex}). Since the core energy is sensitive to
the short-distances properties of the model, different icosadeltahedral 
lattices will have different energies, even for an arbitrarily large number
of particles.

Since most of the triangles in an icosadeltahedral lattice cannot be
equilateral there is no uniquely defined lattice spacing.  An average lattice 
spacing $a$ can, however, be estimated. On the sphere, the distance between two 
nearest-neighbor disclinations is given by $R \gamma$ 
($\gamma=\cos^{-1}(1/\sqrt{5})$). For the $(n,0)$-case we have  the relation 
$R \gamma=n \times a$, and therefore 
\be\label{latt_sp-sphere}
a=\frac{R \gamma}{n} \ .
\ee
Any other sensible way of estimating the lattice spacing, such as the
size of a disclination dipole, should give a value
of the same order. In the following we restrict ourselves to $(n,0)$
lattices for simplicity, but it is easy to generalize 
the formulas to arbitrary $(n,m)$ icosadeltahedral lattices.

  \begin{figure}[htb]
  \epsfxsize=2.5 in \centerline{\epsfbox{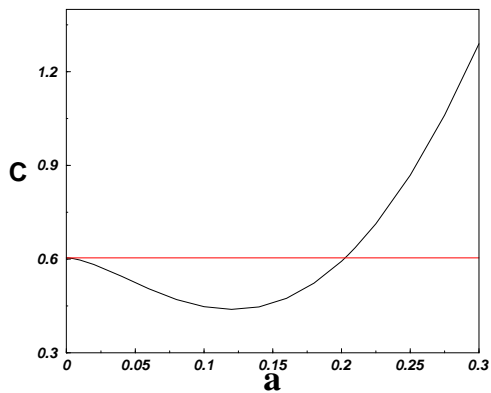}}
  \caption{The $C$ coefficient as a function of the lattice spacing
  for the $C_3$-solution corresponding to a 3 length finite grain boundary.
  The straight line corresponds to the $C$-coefficient for a pure 
  icosahedron.}
  \label{fig__c3sol}
  \end{figure}

From previous sections we know that there will be a range of core
energies for which the icosahedral lattices will be unstable to the
formation of defects.  To visualize this more clearly, let us take the
$C_3$-solution of section~\ref{sub_c3_sol} for the case of finite
grain boundaries having just three defects (a 5-7-5 configuration), 
and plot the $C$-coefficient as a function of the lattice spacing. 
The result is shown in Fig.~\ref{fig__c3sol}.  
For sufficiently large lattice
spacings the $C$ coefficient exceeds that of a single icosahedral
lattice. As the lattice spacing is reduced, there is a critical
particle spacing $a_{*}$, such that the $C$ coefficient of the $C_3$
solution becomes smaller than that of the pure icosahedral
lattice. Restricting ourselves to the $(n,0)$ icosadeltahedral
configurations, we find from Fig.~\ref{fig__c3sol}, Eq.~(\ref{latt_sp-sphere})
and Eq.~(\ref{num_atoms_nm}) 
\be\label{chain_unst}
a_*/R \sim 0.2 \rightarrow n(a_*)=5.5 \rightarrow M(a_*)=305 \ ,
\ee
where we explicitly display the dependence on $a_*$. This result
implies that the pure $(n,0)$ icosadeltahedral lattice is unstable to
the formation of defects for sufficiently small core energies and more
than $316$ particles. Alternatively, Eq.~(\ref{partno}) gives $M=363$, 
consistent with the estimate above. Let us point out that the minimum 
of the energy occurs at lattice spacing $a_c=0.121$ (number of
particles $M_c=1256$). For lattice spacings below $a_c$ the
disclinations will prefer to remain separated by a distance
$a_c$. This is accomplished by stringing six-fold coordinated
particles between defects. Since lower energy configurations may be
formed by allowing interpolating dislocations instead, we regard $M_c$
as the maximum number of particles for which this particular structure
is stable. The particle numbers quoted in the tables of
Sec.~\ref{SECT__SmallCore} should be interpreted as the corresponding
$M_c$. 

This shows that the minimum of the energy is attained for lattice
spacings smaller than those necessary for stability. Finally, in the
limit of vanishing lattice spacing, the $C_3$-solution becomes
equivalent to a pure icosahedron, and both $C$-coefficients merge.

\section{The Thomson problem}\label{SECT__Thomson}

The Thomson problem can be stated as finding the ground state of an
arbitrary number of positive charges interacting through the usual 3d
Coulomb potential, but with the further constraint that these charges
must lie on a sphere. Since this problem falls within the universality
class of our model it serves as a good testing ground.

The Thomson problem has proven to be extremely difficult to analyze
numerically, basically because of the large number of metastable
states.  Early analyses \cite{THOM} showed that the ground state of
the system for small numbers of charges was an icosadeltahedral
lattice.  Since a small number of charges corresponds to a large
particle spacing this follows from our model as well. Some rules were
also conjectured to decide on the true ground state when several
icosadeltahedral lattices were possible for a given number of charges.
These rules could be examined within our model provided we are able to
isolate the dependence of the core energy on the lattice type. 

Subsequent numerical work \cite{NON_ICO_TH} provided convincing
evidence that the ground state, for a sufficiently large number of
charges, does not have icosahedral symmetry.
The critical number of charges for which additional defects arise
seems to be around 400 \cite{ALAR}, which is in agreement with
our results. It is also found that these additional defects first
arrange themselves into finite grain boundaries \cite{ALAR}, as seen
in our model.  For more charges the ground state in the Thomson
problem becomes very complex and the true ground state is not known.
New configurations (one of the simplest being {\em pentagonal
buttons}) appear to be energetically favorable in the early stages of
this limit. This observation \cite{ALAR} is in agreement with our model as well.

In the work of \cite{ALAR} (see also \cite{MOPZ}) it is also observed
that the ground state energy for a large number of charges
seems to converge to the energy that one would obtain in the
unrealizable situation that all the charges are located on equilateral
triangles.  This limit corresponds to the defect density completely
screening out the Gaussian curvature. The defect density
therefore satisfies Eq.~(\ref{zero_screen}), which we proved is the
absolute minimum of our model in the limit of vanishing core energy. 

We think that the comparison of our model with the Thomson problem is
very promising, but requires more detailed investigation to be
addressed in the future.

\section{Discussion and Conclusions}\label{SECT__Conc}

In this paper our first task was to propose and study an effective free energy
for disclination defects in particle arrays constrained to move on the
surface of a two-dimensional sphere. The finite-temperature problem
does not seem to be analytically solvable but we propose a
discretized Laplacian Sine-Gordon model amenable to direct numerical methods. 
The structure of the ground state may, however, be studied
analytically. This structure depends on the ratio of disclination core
energies to the Young's modulus. On the sphere topology demands there be 
a total excess disclinicity charge of twelve. This excess charge can
seed new ground state structures, compared to flat space. For large
core energies (or $R/a \le 36E_{core}/(\pi K_0 a^2)$) the disclinations 
arrange themselves to form an icosahedron. For intermediate core
energies (i.e. $R/a \ge 36E_{core}/(\pi K_0 a^2)$) grain boundaries develop
which terminate freely within the medium. The regime of still lower core
energies, corresponding to $R/a \rightarrow \infty$, was found to be 
surprisingly complex {--} new defect arrangements make their appearance. 

Currently we are actively investigating the regime of small or
vanishing defect core energy, including a detailed comparison of the
predictions of our model with numerical results from the Thomson
problem. A rigorous determination of the ground state for the Thomson
problem is presently computationally prohibitive when the particle 
numbers exceed ${\cal O}(500)$. Our methods enable us to reach 
particle numbers of ${\cal O}(10,000)$ or more with the same
computational effort.  

Finally we believe that the rich symmetry structure underlying 
Eq.~(\ref{zero_screen}) may provide a direct analytic determination of
the exact ground state in the limit of a large number of particles and
further work in this direction is certainly warranted.

\bigskip
\bigskip
\bigskip
\centerline{\bf Acknowledgements}
\bigskip

Our interest in this problem is the result of numerous discussions
with Alar Toomre. One of us (DRN) would like to acknowledge helpful
conversations with F.~Spaepen and B.~I.~Halperin. We are indebted to
S.~Balibar for discussions of the physics of multi-electron bubbles. 
We also acknowledge
use of the software package {\em Geomview} \cite{Geom}. 
The research of MJB and AT was supported by the Department of Energy through Grant
No. DE-FG05-86ER-40272. The research by DRN was supported by the
National Science Foundation through Grant No. DMR97-14725 and through
the Harvard Materials Research Science and Engineering Laboratory via
Grant No. DMR98-09363. Finally MJB would like to acknowledge the
hospitality of Harvard University during a one year stay in which some
of work described in this paper was completed.

\newpage

\appendix

\section{The infinite radius limit}\label{SECT__App__scal}

In this Appendix we discuss in more detail the large $R$ limit of the
energy function Eq.~(\ref{energy_cosb}). It is readily seen from
dimensional analysis and linearity that the $\chi$ function of 
Eq.~(\ref{bi_harm_sol}) scales like $R^2$.

Now consider a single isolated disclination $q_1$ located at point
$P_1$ on the sphere, together with a single dislocation, with Burgers
vector ${\bf b}_2$, located at point $P_2$. From
Eq.~(\ref{energy_cosb}) the total energy is 
\be\label{large_R}
E=\frac{\pi K_0 }{36}\left( q_1^2 R^2 + |{\bf b}_2| q_1 f(P_1,P_2) R +
|{\bf b}_2|^2 \log \left( \frac{R}{2a} \right) \right) \ ,
\ee
where $f(P_1,P_2)$ is a function whose explicit form does not matter for
the present analysis. The quadratic $R$ dependence comes from the 
isolated disclination, the linear $R$ dependence comes from the 
dislocation-disclination interaction and
the logarithmic term comes from the dislocation energy. In the
infinite-radius limit of the sphere we see, therefore, that the various defect
energies scale identically to those in a flat space system of size $R$
\cite{SeuNel,NEL2}. 
The nature of the ground state, however, is dramatically changed.

\section{The biharmonic operator on the sphere}\label{SECT__App__sum}

The evaluation of the inverse biharmonic operator on the sphere is
rather tedious. We outline the steps for a sphere of unit
radius. The simplest approach is to first compute the inverse harmonic operator.
The sum over $m$ is performed as in Eq.~(\ref{bi_harm_Leg}). 
The result involves only Legendre polynomials and is 
\be\label{harm__leg}
\Gamma(x)\equiv \frac{1}{4 \pi \Delta}=\sum^{\infty}_{l=0}(\frac{1}{l+1}+\frac{1}{l})
P_{l}(x) .
\ee
The sums over $l$ may be performed using the identities
\be\label{iden_leg_1}
\sum^{\infty}_{l=0} \frac{1}{l+1} P_l(x)=\int^{1}_0
du\frac{1}{(1-2ux+u^2)^{1/2}}
\ee
and
\be\label{iden_leg_2}
\sum^{\infty}_{l=1} \frac{1}{l} P_l(x)=\lim_{\epsilon \rightarrow 0} 
(\int^{t}_{\epsilon}
du\frac{1}{u(1-2ux+u^2)^{1/2}}-\int^1_{\epsilon} \frac{du}{u}) \ .
\ee
The resultant integrals are readily evaluated, yielding 
\be\label{final_Gamma}
\Gamma(x)=-\log(\frac{1-x}{2})-1 \ .
\ee
The inverse biharmonic operator now follows from the result 
$\Delta\chi(x)=\Gamma(x)$.

\section{The evaluation of $\chi$}\label{SECT__App__Chi}

Performing a trivial integration by parts in Eq.~(\ref{bi_harm_sol}),
we get
\be\label{int_par}
\chi(\beta)=1+t(\log(t)-1)+\int^t_0 dz\frac{z \log z}{1-z} \ ,
\ee
with $t=\frac{1-\cos \beta}{2}$. The last integral 
may be expressed via a change of variables as 
\be\label{new_int}
\int^x_0 dz\frac{z \log z}{1-z}=-\int^{\infty}_u dx \frac{x e^{-x}}{e^x-1}
\ ,
\ee
with $u=ln(1/t)$. Finally one can expand for small $u$ and
large $u$ as
\be\label{order_expansion}
\int^{\infty}_u \frac{x e^{-x}}{e^x-1}=\left\{
{\begin{array}{l} 
\frac{\pi^2}{6}-(u-\frac{u^2}{4}+\sum_{k=1} B_k\frac{u^{2k+1}}{(2k+1)!})
-e^u(u+1) \\
\sum_{n=1}(u+\frac{1}{n+1})\frac{e^{-(n+1)u}}{n+1}
\end{array}}\right. \ ,
\ee
where $B_k$ are the Bernoulli numbers.
These expansions are very useful as they allow a numerical evaluation of
$\chi$ with arbitrary precision with  negligible computational time. 
In fact we save, on average, a factor of 2000 in time compared
to a direct evaluation of the integral using Romberg integration.


\begin{thebibliography}{99}
\bibitem{Remark} We assume for simplicity a finite concentration of
vacancies and interstitials so that the pair can separate via climb
motion. If climb is frozen out kinetically, the dislocations can still
separate via glide motion. The binding potential has an angular
dependence in this case. See Ref.\cite{BruHalZip}.
\bibitem{Comparison} Our results differ from those in Appendix A of
Ref.\cite{DOD}.
\end{thebibliography}
\end{document}